\documentclass[11pt,a4paper]{article}

\usepackage{jheppub}
\usepackage{graphicx}
\usepackage{dcolumn}
\usepackage{bm}
\usepackage{amsmath}
\usepackage{braket}
\usepackage{slashed}    
\usepackage{epstopdf}
\usepackage{placeins}
\usepackage{multirow}
\usepackage{makecell}
\usepackage[shortlabels]{enumitem}
\usepackage{cleveref}
\usepackage{xcolor}
\usepackage[font=small, labelfont=bf, textfont={small,it}]{caption}

\newcommand{\beq}{\begin{eqnarray}}
\newcommand{\eeq}{\end{eqnarray}}
\newcommand{\beqnn}{\begin{eqnarray*}}
\newcommand{\eeqnn}{\end{eqnarray*}}

\newcommand{\Nf}{N_{\scriptscriptstyle{\rm f}}}
\newcommand{\DW}{D_{\scriptscriptstyle{\rm W}}}
\newcommand{\DDW}{D_{\scriptscriptstyle{\rm DW}}}
\newcommand{\DT}{\tilde{D}}
\newcommand{\PL}{P_{\scriptscriptstyle{\rm L}}}
\newcommand{\PR}{P_{\scriptscriptstyle{\rm R}}}
\newcommand{\Nfive}{N_{\scriptscriptstyle{5}}}
\newcommand{\HW}{\mathcal{H}_{\scriptscriptstyle{\rm W}}}

\newcommand{\RMT}{{\scriptscriptstyle{\rm RMT}}}

\newcommand{\R}{{\scriptscriptstyle{\rm R}}}

\newcommand{\ZS}{Z_{\scriptscriptstyle{\rm S}}}

\newcommand{\Zm}{Z_{\scriptscriptstyle{\rm m}}}

\newcommand{\TEK}{{\scriptscriptstyle{\rm TEK}}}
\newcommand{\dd}{\mathrm{d}}
\newcommand{\ee}{\mathrm{e}}
\newcommand{\ii}{\mathrm{i}}
\newcommand{\Tr}{\mathrm{Tr}}
\newcommand{\sign}{\mathrm{sign}}

\newcommand{\SW}{S_{\scriptscriptstyle{\rm W}}}

\newcommand{\SU}{\mathrm{SU}}

\renewcommand{\Im}{\mathrm{Im}}
\renewcommand{\Re}{\mathrm{Re}}

\begin{document}

\title{\centering Universal Features of Chiral Symmetry Breaking in Large-$N$ QCD}

\author[a]{Claudio Bonanno,}
\author[a]{Margarita Garc\'ia P\'erez,}
\author[a,b]{Antonio Gonz\'alez-Arroyo,}
\author[c,d]{\\Ken-Ichi Ishikawa,}
\author[d]{Masanori Okawa,}
\author[e]{Dario Panfalone}

\affiliation[a]{Instituto de F\'isica Te\'orica UAM-CSIC, Calle Nicol\'as Cabrera 13-15,\\Universidad Aut\'onoma de Madrid, Cantoblanco, E-28049 Madrid, Spain}

\affiliation[b]{Departamento de F\'isica Te\'orica, Universidad Aut\'onoma de Madrid,\\M\'odulo 15, Cantoblanco, E-28049 Madrid, Spain}

\affiliation[c]{Core of Research for the Energetic Universe,\\Graduate School of Advanced Science and Engineering,\\Hiroshima University, Higashi-Hiroshima, Hiroshima 739-8526, Japan}

\affiliation[d]{Graduate School of Advanced Science and Engineering, Hiroshima University,\\Higashi-Hiroshima, Hiroshima 739-8526, Japan}

\affiliation[e]{Department of Physics, University of Turin \& INFN, Turin,\\Via Pietro Giuria 1, I-10125 Turin, Italy}

\emailAdd{claudio.bonanno@csic.es}
\emailAdd{margarita.garcia@csic.es}
\emailAdd{antonio.gonzalez-arroyo@uam.es}
\emailAdd{ishikawa@theo.phys.sci.hiroshima-u.ac.jp}
\emailAdd{okawa@hiroshima-u.ac.jp}
\emailAdd{dario.panfalone@unito.it}

\abstract{We investigate the universal features of chiral symmetry breaking in large-$N$ QCD by comparing non-perturbative determinations of the low-lying Dirac spectrum with chiral Random Matrix Theory (RMT) predictions. Our numerical Monte Carlo calculations are based on a chiral lattice discretization of the Dirac operator, and exploit twisted volume reduction to reach $N$ as large as 841. Matching lattice data with RMT analytic results, we are able to extract the large-$N$ chiral condensate, which is compared with a recent determination obtained with non-chiral Wilson quarks from twisted volume-reduced models.}


\maketitle

\section{Introduction}

Spontaneous chiral symmetry breaking is a distinctive feature of QCD which has fundamental phenomenological implications for strong interactions, and tight connections with other prominent non-perturbative properties of the theory, such as $\theta$-dependence and confinement. Due to this crucial role, it has been the subject of a large number of studies in the last decades, aiming at improving our understanding of its microscopic origin and of its universal traits.

In this respect, a staple result is provided by the Banks--Casher relation~\cite{Banks:1979yr}: the realization of chiral symmetry in QCD puts strong constraints on the properties of the low-lying Dirac spectrum. Building on this fundamental finding, it has been argued~\cite{Leutwyler:1992yt,Shuryak:1992pi,Verbaarschot:1993pm,Verbaarschot:1994qf,Nishigaki:1998is} that the eigenvalues of the Dirac operator follow the same universal probability distributions as those of an ensemble of random matrices, where the whole dependence on the microscopic details of the   full theory is encoded in a single parameter, identified with the chiral condensate $\Sigma=-\braket{\overline{\psi}\psi}$. In the fundamental representation and when $N$, the number of colors, is larger than 2, the universality class is described by the so-called chiral unitary Random Matrix Theory (RMT). By now, non-perturbative lattice calculations have clearly established this correspondence in QCD from first-principles, both in the chirally-broken phase~\cite{Edwards:1999ra,Damgaard:1999tk,Bietenholz:2003mi,Giusti:2003gf,Wennekers:2005wa,Bernardoni:2008ei,Hasenfratz:2008ce,Bernardoni:2010nf,Deuzeman:2011dh,Catillo:2017qbz} and in the chirally-restored phase sufficiently above the Anderson mobility edge~\cite{Kovacs:2012zq,Giordano:2013taa,Kovacs:2017uiz,Vig:2020pgq,Giordano:2021qav}, by comparing numerical results for the Dirac spectrum with analytic RMT predictions (see also Refs.~\cite{Verbaarschot:2000dy,Damgaard:2001ep,Verbaarschot:2009jz,Damgaard:2011gc} for reviews and further references).

Apart from standard QCD, chiral symmetry breaking plays a fundamental theoretical and phenomenological role also in other QCD-like gauge theories, such as large-$N$ QCD in the 't Hooft limit, obtained when the number of colors $N$ is taken to infinity $N\to \infty$ at a fixed 't Hooft coupling $\lambda=g^2N$ and for a fixed number of quark flavors $\Nf$ (i.e., $\Nf/N\to 0$). For this reason, in the last two decades chiral symmetry breaking has been extensively studied in this regime on the lattice. However, most of these studies addressed the determination of the chiral condensate either through the Banks--Casher relation, or through the quark mass dependence of the pion mass~\cite{Narayanan:2005gh,DelDebbio:2007wk,Bali:2013kia,DeGrand:2016pur,Hernandez:2019qed,DeGrand:2023hzz,Bonanno:2023ypf,Bonanno:2025hzr}, while the investigation of the realization of the universal chiral RMT behavior of the large-$N$ Dirac spectrum has only been addressed in a few seminal papers~\cite{Narayanan:2004cp,Gonzalez-Arroyo:2005dgf} (see also~\cite{Hanada:2013ota}). The goal of the present paper is to fill this gap and address the universal features of chiral symmetry breaking in large-$N$ QCD via state-of-the-art lattice calculations. To do so, we will consider the Twisted Eguchi--Kawai (TEK) model~\cite{GONZALEZARROYO1983174,PhysRevD.27.2397,Gonzalez-Arroyo:2010omx}.

The TEK model is an effective framework to study lattice gauge theories in the large-$N$ limit by exploiting a fundamental property of large-$N$ gauge theories: the so-called large-$N$ volume reduction~\cite{PhysRevLett.48.1063,BHANOT198247,Gross:1982at,GONZALEZARROYO1983174,PhysRevD.27.2397,Aldazabal:1983ec,Kiskis:2002gr,Narayanan:2003fc,Kovtun:2007py,Unsal:2008ch,Gonzalez-Arroyo:2010omx,Neuberger:2020wpx}. In their pioneering paper, Eguchi and Kawai~\cite{PhysRevLett.48.1063} showed that, in the large-$N$ limit, the Yang--Mills lattice theory enjoys a dynamical equivalence between space-time and color degrees of freedom, leading the model to become insensitive to finite-volume effects when $N\to\infty$. Taking this idea to the extreme, it allows to study large-$N$ gauge theories in the thermodynamic limit even on a volume-reduced 1-site lattice~\cite{Gonzalez-Arroyo:1983cyv,Das:1984jh,Das:1985xc,Kiskis:2009rf,Hietanen:2009ex,Hietanen:2010fx,Bringoltz:2011by,Hietanen:2012ma,Gonzalez-Arroyo:2012euf,Gonzalez-Arroyo:2013bta,Lohmayer:2013spa,Gonzalez-Arroyo:2014dua,GarciaPerez:2014azn,GarciaPerez:2015rda,Perez:2015ssa,Gonzalez-Arroyo:2015bya,Perez:2017jyq,GarciaPerez:2020gnf,Perez:2020vbn,Butti:2023hfp,Bonanno:2023ypf,Butti:2022sgy,Bonanno:2024bqg,Bonanno:2024onr}. This has the advantage of dramatically reducing the computational burden related to the space-time degrees of freedom, and allows to reach very large values of $N\sim\mathcal{O}(10^2-10^3)$ in feasible simulations. Thus, unlike standard approaches to study large-$N$ gauge theories --- where the $N=\infty$ limit is typically approached from $N\lesssim10$ values via extrapolation~\cite{DelDebbio:2001sj,Lucini:2001ej,DelDebbio:2002xa,Lucini:2004my,DelDebbio:2006yuf,Vicari:2008jw,Allton:2008ty,Lucini:2010nv,Lucini:2012gg,Bali:2013kia,Bonati:2016tvi,Ce:2016awn,Bennett:2020hqd,Bonanno:2020hht,Athenodorou:2021qvs,Bennett:2022gdz,Bonanno:2022yjr,Bonanno:2024ggk,Athenodorou:2024loq,Sharifian:2025fyl, DeGrand:2016pur,Hernandez:2019qed,DeGrand:2020utq,Hernandez:2020tbc,DeGrand:2021zjw,Baeza-Ballesteros:2022azb,DeGrand:2023hzz,DeGrand:2024lvp,DeGrand:2024frm,Baeza-Ballesteros:2025iee,Butti:2025rnu} --- the TEK model practically allows to work directly at $N=\infty$. Large-$N$ volume reduction works provided that center symmetry is unbroken. Spontaneous center symmetry breaking that would occur when reducing the lattice size due to the well-known deconfinement transition~\cite{Fingberg:1992ju,Beinlich:1997ia,Campostrini:1998zd,Lucini:2001ej,Lucini:2002ku,Lucini:2003zr,Lucini:2004yh,Lucini:2005vg,Lucini:2012wq,Borsanyi:2022xml,Lucini:2023irm,Cohen:2023hbq} is avoided in the TEK approach by imposing twisted boundary conditions on the gauge fields in all directions, hence the name of the model.\footnote{This is not the only strategy that has been applied to enforce center symmetry on a small box and achieve large-$N$ volume independence see, e.g., Refs.~\cite{BHANOT198247,Narayanan:2003fc,Kovtun:2007py,Unsal:2008ch}.}

In order to clarify how our method compares with more traditional ones, it is useful to recall that the nature of finite-$N$ corrections is different in the TEK model
than in standard approaches. The model at finite $N$ is actually
related to the corresponding non-commutative field theory
models~\cite{Gonzalez-Arroyo:1983zev,Connes:1987ue,Rieffel:1998vs,Ambjorn:2000cs,Douglas:2001ba,Alvarez-Gaume:2001czc}. The form of these corrections is already clear from the results in the original paper~\cite{PhysRevD.27.2397}. The planar diagram contribution appears as the one formulated in a finite lattice of size $(\sqrt{N})^4$. Thus, as it will be done in this paper, the size of the lattice in physical units has to be taken large enough to make these corrections small for the observable in question. This, however, is not different from the minimal volume requirements in more
conventional approaches. Indeed, one can take advantage of the latter
studies to quantify the importance of these corrections.
In addition, finite-$N$ values in the TEK model also affect the suppression of non-planar contributions. This effect has been studied extensively in more recent times starting in Ref.~\cite{Gonzalez-Arroyo:2010omx}. It turns out that the choice of the flux parameter $k$, appearing in the twist factor implementing twisted boundary conditions in the TEK action (see Sec.~\ref{sec:setup}), plays a crucial role in keeping these corrections small. An appropriate $N$-dependent choice $k(N)$ gives rise to corrections of order $1/N^2$ with a small prefactor.
This, given the large values of $N$ achievable in the model, makes
these corrections negligibly small. All these effects have been studied 
in detail in the recent literature both perturbatively~\cite{Perez:2017jyq,Gonzalez-Arroyo:2019zfm} and non-perturbatively~\cite{Gonzalez-Arroyo:2014dua}. As a matter of fact, the TEK approach has been extensively used in lattice simulations in the last decade, and has allowed to achieve significant progress in the study of several large-$N$ gauge theories, such as large-$N$ QCD~\cite{Gonzalez-Arroyo:2012euf,GarciaPerez:2014azn,Gonzalez-Arroyo:2015bya,Perez:2020vbn,Bonanno:2023ypf,Butti:2023hfp,Bonanno:2025hzr}, large-$N$ SUSY Yang--Mills~\cite{Butti:2022sgy,Bonanno:2024bqg,Bonanno:2024onr}, and large-$N$ adjoint QCD~\cite{Gonzalez-Arroyo:2013bta,GarciaPerez:2015rda,Hamada:2025whg}.

In order to compare RMT predictions with our large-$N$ non-perturbative determinations of the low-lying Dirac eigenvalues, it is of the utmost importance to preserve chiral symmetry even at finite lattice spacing. It is thus imperative to use a lattice formulation of the Dirac operator that satisfies the Ginsparg--Wilson relation~\cite{Ginsparg:1981bj}. For this reason, we implemented an overlap discretization~\cite{Neuberger:1997fp,Neuberger:1998wv,Luscher:1998pqa} of chiral quarks within our TEK model via the so-called ``truncated overlap'' approach~\cite{Kaplan:1992bt,Shamir:1993zy,Furman:1994ky,Borici:1999zw,Edwards:2000qv,Ishikawa:2013rxa}. This approach is presented here for the first time, as previous TEK studies have all employed non-chiral Wilson fermions.

Our study of the universal traits of the Dirac spectrum will be articulated in two main parts. First, we will compare our results with scale-invariant RMT predictions in order to establish the universal behavior of Dirac eigenvalues in a parameter-agnostic fashion. Then, we will match our results to parameter-dependent RMT analytic results in order to extract the large-$N$ quark condensate $\Sigma/N\sim\mathcal{O}(N^0)$. This will be important for two reasons: first, to verify that different lattice--RMT matching prescriptions lead to the same value of the condensate, and second, to see how this chiral determination of $\Sigma/N$ compares with the latest TEK one obtained in~\cite{Bonanno:2025hzr} with non-chiral Wilson quarks.

This paper is organized as follows: in Sec.~\ref{sec:RMT} we provide a compact summary of a few RMT predictions that we aim at checking; in Sec.~\ref{sec:setup} we discuss our lattice setup, focusing in particular on our TEK implementation of the chiral lattice Dirac operator; in Sec.~\ref{sec:res} we discuss our numerical results, focusing on the comparison of lattice data with RMT predictions and on the extraction of the chiral condensate; finally, in Sec.~\ref{sec:conclu} we draw our conclusions and discuss future outlooks of this study.

\section{Random Matrix Theory predictions}\label{sec:RMT}
The relevant universality class to describe large-$N$ QCD is the chiral unitary RMT. Within this effective model, it is possible to derive analytic predictions for the probability distributions $p_k(z_k)$ of the $k^{\rm th}$ eigenvalue $z_k$ of the random matrix ensemble for any fixed topological sector $Q$~\cite{Damgaard:2000ah} (in this discussion the spectrum is assumed to be ordered increasingly, and any degeneracy in the spectrum is assumed to be factored out). As an example, in the $Q=0$ sector, the distributions of the two lowest eigenvalues $z_1$ and $z_2$ and of their ratio $r=z_1/z_2$ are given by~\cite{Damgaard:2000ah,Narayanan:2004cp}:
\beq
\label{eq:p_1_prediction}
p_1(z_1) &=& \frac{1}{2}z_1 \ee^{-\frac{z_1^2}{4}}, \\
\label{eq:p_2_prediction}
p_2(z_2) &=& \frac{1}{4} \ee^{-\frac{z_2^2}{4}} z_2  \int_0^{z_2} \dd u \, u \, \left[I_2(u)^2 - I_1(u)I_3(u)\right], \\
\label{eq:p_r_prediction}
p(r) &=& \frac{1}{4} \frac{r}{(1 - r^2)^2} \int_0^\infty \dd u \, \ee^{- \frac{u^2}{4(1 - r^2)}} \, u^3 \left[ I_2(u)^2 - I_1(u) I_3(u) \right],
\eeq
with $I_n(x)$ the $n^{\rm th}$ modified Bessel function of the first kind,
\beq
I_n(x) = \sum_{m \, = \, 0}^{\infty}\frac{1}{(m+n)! \, m!} \left(\frac{x}{2}\right)^{2m+n} = \int_0^{2 \pi} \frac{\dd\theta}{2 \pi} \, \ee^{\ii \theta n} \,   \ee^{x\cos(\theta)} \, .
\eeq

In the thermodynamic limit $V\to\infty$, one expects the following identification to hold, topological sector by topological sector, for the eigenvalues of the massless Dirac operator:
\beq\label{eq:QCD_RMT_matching}
z_k = \Sigma V \lambda_k, \qquad \slashed{D}v_{\lambda_k} = \ii \lambda_k v_{\lambda_k}, \qquad \lambda_k \in \mathbb{R}.
\eeq
Thus, for asymptotically large volumes, Eq.~\eqref{eq:QCD_RMT_matching} is expected to bridge the RMT eigenvalues $z_k$ and the corresponding Dirac eigenvalues $\lambda_k$ for all $k$ via a single parameter, the quark condensate $\Sigma$. Since on the lattice one typically computes the spectrum of the massless chiral lattice Dirac operator in a finite volume~\cite{Narayanan:2004cp,Gonzalez-Arroyo:2005dgf,Giusti:2003gf,Wennekers:2005wa,Bernardoni:2008ei}, one expects to see deviations from chiral RMT predictions, which should become smaller and smaller and occur for higher and higher $k$ as $V\to\infty$. Such deviations typically manifest in the fact that, assuming Eq.~\eqref{eq:QCD_RMT_matching}, different choices of $k$ lead to different results for $\Sigma$.

For this reason, our check of the universal features of the large-$N$ Dirac spectrum will be subdivided into two steps. First, we will monitor deviations of lattice Dirac eigenvalues with respect to RMT predictions in a parameter-free, scale-invariant way by computing the deviation from 1 of the following ratios as a function of the volume:
\beq \label{eq:ratio_rmt}
R_{k_1,k_2} \equiv \frac{\left\langle \lambda_{k_1}\right\rangle}{\left\langle\lambda_{k_2}\right\rangle} \times \left(\frac{\left\langle z_{k_1}\right\rangle_{\RMT}}{\left\langle z_{k_2}\right\rangle_{\RMT}}\right)^{-1}.
\eeq
Here $\braket{\mathcal{O}}$ and $\braket{\mathcal{O}}_{\RMT}$ denote respectively the expectation values taken in the full theory and in the RMT effective model. Given that in these eigenvalue ratios the factor of $\Sigma V$ drops and no free parameter can be adjusted, this comparison constitutes a very strong test of the universal traits of the spectrum.

Then, once the correct volume regime has been established, we will move to parameter-dependent RMT predictions, and we will test them by matching lattice data and analytic results to extract the quark condensate assuming the relation~\eqref{eq:QCD_RMT_matching}. This can be done either by computing the full probability distribution of the first few eigenvalues and fitting it to RMT predictions, such as those in Eqs.~\eqref{eq:p_1_prediction}--\eqref{eq:p_r_prediction}, or by computing the following ratios:
\beq\label{eq:RMT_condensate}
\Sigma V = \frac{\left\langle z_k\right\rangle_{\RMT}}{\left\langle\lambda_k\right\rangle}.
\eeq
These two routes should of course lead to compatible estimates. Since all RMT predictions are described by the same single parameter $\Sigma$, we will check that different matching prescriptions --- i.e., different choices of $k$ --- to extract the chiral condensate give compatible result for the condensate when the corresponding $\lambda_k$ eigenvalues are in the RMT regime. Moreover, we will compare these determinations of $\Sigma$ with the one found in~\cite{Bonanno:2025hzr} for non-chiral Wilson quarks.

\section{Numerical setup}\label{sec:setup}

This section provides an overview of our lattice discretization for the gluon and the quark sectors, with particular emphasis on the adopted chiral formulation of the Dirac operator, which is implemented for the TEK model in this study for the first time.

\subsection{Lattice action}

In the 't Hooft large-$N$ limit quarks are quenched, i.e., non-dynamical. From the point of view of lattice simulations, this means that we will draw gluonic configurations from the Monte Carlo according to the pure-gauge action, ignoring the fermion determinant. Then, these gluon fields will constitute the background of the chiral lattice Dirac operator (defined in the next section) which will be employed for spectrum computations.

The TEK partition function of the $d=4$ single-site TEK model describing the gluonic sector reads:
\beq\label{eq:part_func}
Z_\TEK \equiv \int [\dd U]\, \ee^{-\SW[U]}, \qquad [\dd U] \equiv \prod_{\mu\,=\,1}^{d}\left[\dd U_\mu\right],
\eeq
with $[\dd U]$ the $\SU(N)$ invariant Haar measure and
\beq\label{eq:TEK_Wilson_action}
\SW[U] = -N b \sum_{\mu\,=\,1}^{d}\sum_{\nu \,\ne\, \mu} z_{\nu\mu} \Tr\left\{U_\mu U_\nu U_\mu^\dagger U_\nu^\dagger\right\},
\eeq
the TEK Wilson plaquette action. The Monte Carlo algorithm employed for the sampling of this functional integral is described in detail in Ref.~\cite{Perez:2015ssa}. In Eq.~\eqref{eq:TEK_Wilson_action} $b=1/(Ng^2)$ is the inverse bare 't Hooft coupling, $U_\mu$ are the $d=4$ $\SU(N)$ gauge link matrices describing lattice gluon fields, and $z_{\nu \mu}$ is the twist factor used to implement twisted boundary conditions. It is chosen to be a $N^{\text{th}}$-root of unity:
\beq\label{eq:twist_def}
z_{\nu\mu} = z_{\mu\nu}^* = \exp\left\{\frac{2\pi \ii}{N} n_{\nu\mu} \right\}.
\eeq
There are several possible choices of the twist factor; here, we will adopt the so-called \emph{symmetric} twist. This means that $N$ is taken to be a perfect square $N=L^2$, and that the integer-valued anti-symmetric twist tensor $n_{\nu \mu}$ is taken to be:
\beq
n_{\nu \mu} = - n_{\mu \nu} = k(L) L, \quad \text{($\nu>\mu$)},
\eeq
with $k(L)$ a co-prime integer with $L$. In the end, thus:
\beq
z_{\nu\mu} = \exp\left\{2\pi \ii\frac{k(L)}{L} \varepsilon_{\nu\mu} \right\}, \qquad \varepsilon_{\nu\mu} = - \varepsilon_{\mu\nu}, \qquad \varepsilon_{\nu\mu} = 1 \quad \text{($\nu>\mu$)}.
\eeq
As explained in Refs.~\cite{Gonzalez-Arroyo:2010omx,Chamizo:2016msz,GarciaPerez:2018fkj}, the flux parameter $k(L)$ and its inverse (mod $L$) $\bar{k}(L)$ have to be scaled with $L$ to keep $|k|/L$ and $|\bar{k}|/L$ bounded from below (by $\sim 0.1$) in order to avoid center-symmetry breaking~\cite{Ishikawa:2003,Bietenholz:2006cz,Teper:2006sp,Azeyanagi:2007su} that would invalidate one of the assumptions behind large-$N$ volume reduction. In addition, it has been shown in perturbation theory that suitable choices of $k,\bar{k}$ help in reducing non-planar finite-$N$ corrections~\cite{Perez:2017jyq,Bribian:2019ybc}.

As explained in the introduction, even if the TEK model is formulated on a reduced 1-point box, this does not mean at all that our fields propagate in a torus of vanishing volume. Indeed, by virtue of large-$N$ volume independence and of the adopted twisted boundary conditions, physical excitations actually propagate on an extended torus with an effective physical size given by:
\beq
\ell = a L = a \sqrt{N},
\eeq
with $a(b)$ the lattice spacing. This means that in our setup, at fixed coupling $b$, the thermodynamic limit is achieved when $N\to\infty$. It is in this limit that we expect RMT predictions to hold.

\subsection{Lattice chiral Dirac operator}

The non-chiral TEK Wilson discretization of the Dirac operator, first discussed in Ref.~\cite{Gonzalez-Arroyo:2015bya}, is given by the following $N^2\times N^2$ matrix (where $N^2$ is the size of the effective volume in lattice units in the TEK approach):
\beq\label{eq:wilson_dirac}
\DW(m_{\scriptscriptstyle{\rm W}}) &=& 4+m_{\scriptscriptstyle{\rm W}} - \frac{1}{2} \sum_{\mu \, =\, 1}^{d}\left[(1+\gamma_\mu) \otimes \mathcal{W}_\mu + (1-\gamma_\mu)\otimes \mathcal{W}_\mu^\dagger \right],\\
\nonumber\\[-1em]
\mathcal{W}_\mu &=& U_\mu \otimes \Gamma_\mu^*, \qquad \Gamma_\mu \Gamma_\nu = z_{\nu\mu}^* \Gamma_\nu \Gamma_\mu,
\eeq
with $\Gamma_\mu$ the twist eaters~\cite{Gonzalez-Arroyo:1997ugn}, $\SU(N)$ matrices satisfying $\Gamma_\mu \Gamma_\nu = z_{\nu\mu}^* \Gamma_\nu \Gamma_\mu$, with $z_{\nu\mu}$ the same twist factor appearing in~\eqref{eq:twist_def}. To derive this expression, one lets quarks live on an extended periodic lattice with periodicity $\sqrt{N}$ (recall that $z_{\nu\mu}^{\sqrt{N}}=1$), and let them interact with a periodic potential obtained by replicating the 1-site gauge fields $\sqrt{N}$ times. To some extent, this is reminiscent of the Bloch description of electrons in a crystal.

Building on this non-chiral definition, we can now proceed to define a chiral lattice Dirac operator. For this purpose, we have implemented a chiral Dirac operator using the so-called ``truncated overlap'' formulation~\cite{Borici:1999zw,Edwards:2000qv,Kikukawa:2000ac,Ishikawa:2013rxa}, whose definition will be summarized in the following. The starting point is the $5d$ M\"obius Domain Wall (DW) Dirac operator~\cite{Brower:2005qw} (here $\Nfive$ is the size of the $5^{\rm th}$ dimension):
\beq\label{eq:domain_wall_def}
\DDW(\hat{m},M,\Nfive) = \DW(-M) X(\hat{m},\Nfive) + Y(\hat{m},\Nfive),
\eeq
with $\hat{m}$ the bare quark mass in lattice units. The operator $\DW(-M)$ is the TEK Wilson Dirac operator~\cite{Gonzalez-Arroyo:2015bya} earlier discussed with a negative kernel mass $M = 1+s$ $(0\le s < 1)$, and where we have replaced the standard links $U_\mu$ with stout smeared~\cite{Morningstar:2003gk} links $\overline{U}_\mu$. Instead, the $X(\hat{m},\Nfive)$ and $Y(\hat{m},\Nfive)$ operators appearing in Eq.~\eqref{eq:domain_wall_def} are matrices acting on the $5^{\rm th}$ dimension, whose expressions are:
\beq
X(\hat{m},\Nfive) &=& B + C M_5(\hat{m})\\
\nonumber\\[-1em]
Y(\hat{m},\Nfive) &=& 1 + M_5(\hat{m})\\
\nonumber\\[-1em]
M_5^{(s,t)}(\hat{m}) &=& \left(\PL \delta_{s+1,t} + \PR \delta_{s-1,t}\right) - \hat{m}\left(\PL \delta_{s,\Nfive}\delta_{1,t} + \PR \delta_{s,1} \delta_{t,\Nfive}\right)
\eeq
with $\PL = (1+\gamma_5)/2$, $\PR = (1-\gamma_5)/2$ the chiral projectors, and with
\beq
B^{(s,t)} &=& b_s \, \delta_{s,t}\,,\\
\nonumber\\[-1em]
C^{(s,t)} &=& c_s \, \delta_{s,t}\,,
\eeq
diagonal matrices in the $5^{\rm th}$ dimension. The choice of the vectors $b_s$ and $c_s$ will be discussed later in this section. From $\DDW$, the four-dimensional truncated overlap operator $\DT$ is built via the following projection:
\beq
\DT(\hat{m},M,\Nfive) = \left[\mathcal{P}\DDW^{-1}(M,\hat{m}=1,\Nfive)\DDW(M,\hat{m},\Nfive)\mathcal{P}\right]_{1,1},
\eeq
with $\mathcal{P} = \PL\delta_{s,t} + \PR(\delta_{s+1,t}+\delta_{s,1}\delta_{t,\Nfive})$. This yields the following expression for the truncated overlap~\cite{Borici:1999zw,Kikukawa:2000ac,Edwards:2000qv,Brower:2005qw} that we will use in our study:
\beq
\DT(\hat{m},M,\Nfive) = \frac{1+\hat{m}}{2} \, 1 + \frac{1-\hat{m}}{2} \gamma_5 \, S(M,\Nfive).
\eeq
In this expression, the function $S(M,\Nfive)$ is given by:
\beq
S(M,\Nfive) = \frac{\prod_{s \, = \, 1}^{\Nfive}\left[1+\HW^{(s)}(M)\right]-\prod_{s \, = \, 1}^{\Nfive}\left[1-\HW^{(s)}(M)\right]}{\prod_{s \, = \, 1}^{\Nfive}\left[1+\HW^{(s)}(M)\right]+\prod_{s \, = \, 1}^{\Nfive}\left[1-\HW^{(s)}(M)\right]},
\eeq
with
\beq
\HW^{(s)}(M) = (b_s+c_s)\gamma_5 \DW(-M)\left[(b_s-c_s)\DW(-M)+2\right]^{-1}.
\eeq
The choice of the vectors $b_s$ and $c_s$ is arbitrary as long as $S(M,\Nfive)$ reproduces the sign function of overlap fermions in the limit $\Nfive\to\infty$. Different choices correspond to different lattice formulations, including Shamir-, Boriçi-, and Chiu-type realizations of domain wall fermions~\cite{Shamir:1993zy,Furman:1994ky,Borici:1999zw,Chiu:2002ir}, all agreeing in the continuum limit. In this study we chose $b_s = 1$ and $c_s = 0$, corresponding to the Shamir operator. With this choice, we obtain:
\beq
S(M,\Nfive) &=& \frac{\left[1+\HW(M)\right]^{\Nfive}-\left[1-\HW(M)\right]^{\Nfive}}{\left[1+\HW(M)\right]^{\Nfive}+\left[1-\HW(M)\right]^{\Nfive}},
\eeq
with $\HW(M)$ the ($\gamma_5$-Hermitian) kernel
\beq\label{eq:kernel}
\HW(M) &=& \gamma_5 \DW(-M) \left[2+\DW(-M)\right]^{-1}.
\eeq
In the limit $\Nfive \to \infty$, when the size of the $5^{\rm th}$ dimension goes to infinity, we recognize that $S(M,\Nfive)$ tends to the sign of the kernel $\HW(M)$:
\beq
\lim_{a\,\to\,\infty}\frac{(1+x)^a-(1-x)^a}{(1+x)^a+(1-x)^a} = \sign(x).
\eeq
Therefore, in the limit $\Nfive \to \infty$, the truncated overlap tends to the actual overlap operator:
\beq
\lim_{\Nfive \, \to \, \infty} \DT(\hat{m},M,\Nfive) = D(\hat{m},M)\ ,
\eeq
which takes the form
\beq\label{eq:overlap}
D(\hat{m},M) = \frac{1+\hat{m}}{2} \, 1 + \frac{1-\hat{m}}{2} \, V(M) = (1-\hat{m})D_0(M) + \hat{m},
\eeq
with $V(M) = \gamma_5 \, \sign\left[\HW(M)\right]$, and $D_0(M)$ the massless overlap operator,
\beq\label{eq:overlap_massless}
D_0(M) = \frac{1+V(M)}{2}.
\eeq

In order to restore physical units in the lattice operator $D$ and in the lattice mass $\hat{m}$ and connect them to their physical counterparts, one needs to perform the following rescalings:
\beq\label{eq:M_rescaling_phys_units}
\slashed{D} &=& \frac{M(2-M)}{a} D_0(M),\\
m &=& \frac{M(2-M)}{a} \hat{m},
\eeq
where $m$ is the \emph{bare} quark mass in physical units. The prefactor $M(2-M)$, which stems from the choice of the kernel in Eq.~\eqref{eq:kernel}, was worked out analytically by requiring that the spectrum of the massless operator in Eq.~\eqref{eq:M_rescaling_phys_units} in the free case was equal to $\slashed{p}$ in momentum space. More details can be found in the Appendix (see also Ref.~\cite{Aoki:2001su}). Clearly, the same normalization applies when restoring physical units in the eigenvalues of the massless overlap Dirac operator:
\beq
D_0(M) u_{\hat{\lambda}} &=& \hat{\lambda}(M) u_{\hat{\lambda}},\\
\hat{\lambda}(M) &=& \frac{1}{2}\left[1+\ee^{\ii\theta(M)}\right],\\
\lambda_{\rm o} &=& \frac{M(2-M)}{a} \hat{\lambda}(M). \label{eq:scale}
\eeq
Apart from this necessary overall rescaling, $M$ is expected to play no other role, and any possible weak $M$-dependence observed in a physical quantity is just a lattice artifact vanishing in the continuum limit, see, e.g., Ref.~\cite{DelDebbio:2004ns}.

\section{Numerical results}\label{sec:res}

This section is devoted to present our non-perturbative lattice results for the low-lying chiral Dirac spectrum, and to discuss the comparison with RMT predictions. As already anticipated, we will conduct such comparison both in a parameter-free and in a parameter-dependent fashion. The latter approach will also allow the extraction of the large-$N$ chiral condensate $\Sigma/N\sim\mathcal{O}(N^0)$, whose value will be compared with the latest large-$N$ TEK determination with non-chiral Wilson quarks~\cite{Bonanno:2025hzr}. Simulation parameters are summarized in Tab.~\ref{tab:params}, along with the value of the string tension $\sigma$ from the TEK model~\cite{Gonzalez-Arroyo:2012euf}, used for scale setting.

\begin{table}[!t]
\begin{center}
\begin{tabular}{|c|c|c|c|c|c|c|c|c|}
\hline
$b$ & $N$ & $L=\sqrt{N}$ & $k$ & $k/L$ & $\bar{k}$ & $\bar{k}/L$ & $a\sqrt{\sigma}$ & $n_{\rm conf}$\\
\hline
\hline
0.355 & 529 & 23 & 7 & 0.304 & 10 & 0.435 & 0.2410(30) &400\\ 
\hline
\hline
\multirow{4}{*}{0.360} & 289 & 17 & 5 & 0.294 & 7 & 0.412 & \multirow{4}{*}{0.2058(25)} &326\\
& 361 & 19 & 7 & 0.368 & 11 & 0.579 &&189\\
& 529 & 23 & 7 & 0.304 & 10 & 0.435 &&222\\
& 841 & 29 & 9 & 0.310 & 13 & 0.448 &&210\\
\hline
\end{tabular}
\end{center}
\caption{Values of the number of colors $N$, of the effective size $L=\sqrt{N}$, and of the parameters $k,\bar{k}$ [where $\bar{k}k=1~(\mathrm{mod}~L)$] used in this study. The string tension $\sigma$ was computed in the TEK model in~\cite{Gonzalez-Arroyo:2012euf}. Finally, $n_{\rm conf}$ is the number of statistically-independent gauge configurations employed.}
\label{tab:params}
\end{table}

\subsection{Checking the chiral properties of the lattice Dirac operator}

We start our discussion by examining the chiral properties of the truncated overlap operator. Indeed, one should check that the adopted $\Nfive$ is large enough to have sufficiently small chirality violations (i.e., a negligible residual mass). These are monitored through the figure of merit:
\beq\label{eq:GW_violation}
\Delta = \frac{1}{\vert z \vert} \Bigg\vert \left( \{\DT,\gamma_5\} - 2\DT\gamma_5\DT \right) z \Bigg\vert,
\eeq
which quantifies the violation of the Ginsparg--Wilson relation.

The violation parameter $\Delta$ was computed stochastically using 3 random $\mathbb{Z}_4$-sources $z$ with color and Dirac indices. After a few tests, we concluded that, in order to achieve satisfactory small values of $\Delta$, smaller values of $\Nfive$ were sufficient when the gauge links entering the truncated overlap Dirac operator were smeared. As an example, $\Delta \sim 10^{-8}$ was achieved with $\Nfive=60$ and no smearing, while the same result was obtained choosing $\Nfive=24$ and $n_s = 5$ stout-smearing steps (with isotropic stout parameter $\rho=0.1$~\cite{Morningstar:2003gk}). The latter choice of smearing parameters corresponds to a smearing radius~\cite{Alexandrou:2017hqw}:
\beq
\frac{R_s}{a} = \sqrt{8 \rho \, n_s} = 2.
\eeq
This is equal in magnitude to the choice adopted in other stout-smeared definitions of the lattice Dirac operator employed in recent state-of-the-art lattice QCD simulations. For example, the BMW collaboration recently employed a stout-smeared definition of the staggered Dirac operator with $n_s=4$ and $\rho=0.125$ --- corresponding to $R_s/a=2$ --- in their recent lattice investigations of the hadronic vacuum polarization contribution to the muon anomalous magnetic moment~\cite{Borsanyi:2020mff,Boccaletti:2024guq}.\\
The choice of using stout-smeared links was also useful to reduce the computational burden and make $N=841$ calculations feasible. Indeed, at fixed $N$, $\Nfive$, and number of computed eigenvalues, we found that spectra calculations involving the stout-smeared operator were $\sim 3$ times less expensive with respect to the non-smeared case, on top of the $\sim 2.5$ reduction of $\Nfive$ at fixed $\Delta$ achieved using stout smearing, which decreases the computational effort by the same factor. Thus, all our results have been obtained for $\Nfive=24$ and stout-smeared gauge links with $n_s=5$ and $\rho=0.1$, yielding $\Delta \lesssim 10^{-8}$ for all explored values of $N$ and $b$.

\begin{figure}[!t]
\centering
\includegraphics[scale=0.33]{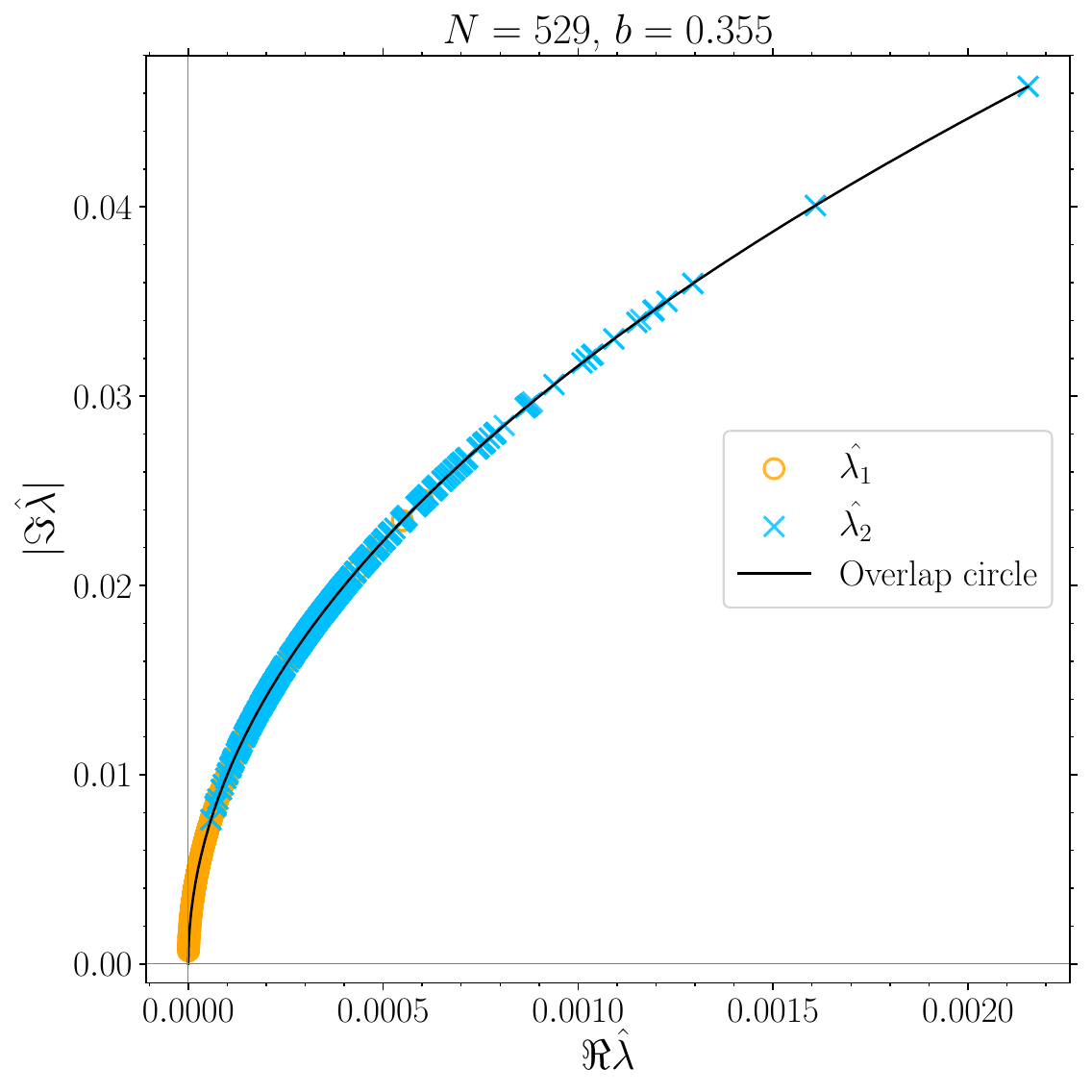}
\includegraphics[scale=0.33]{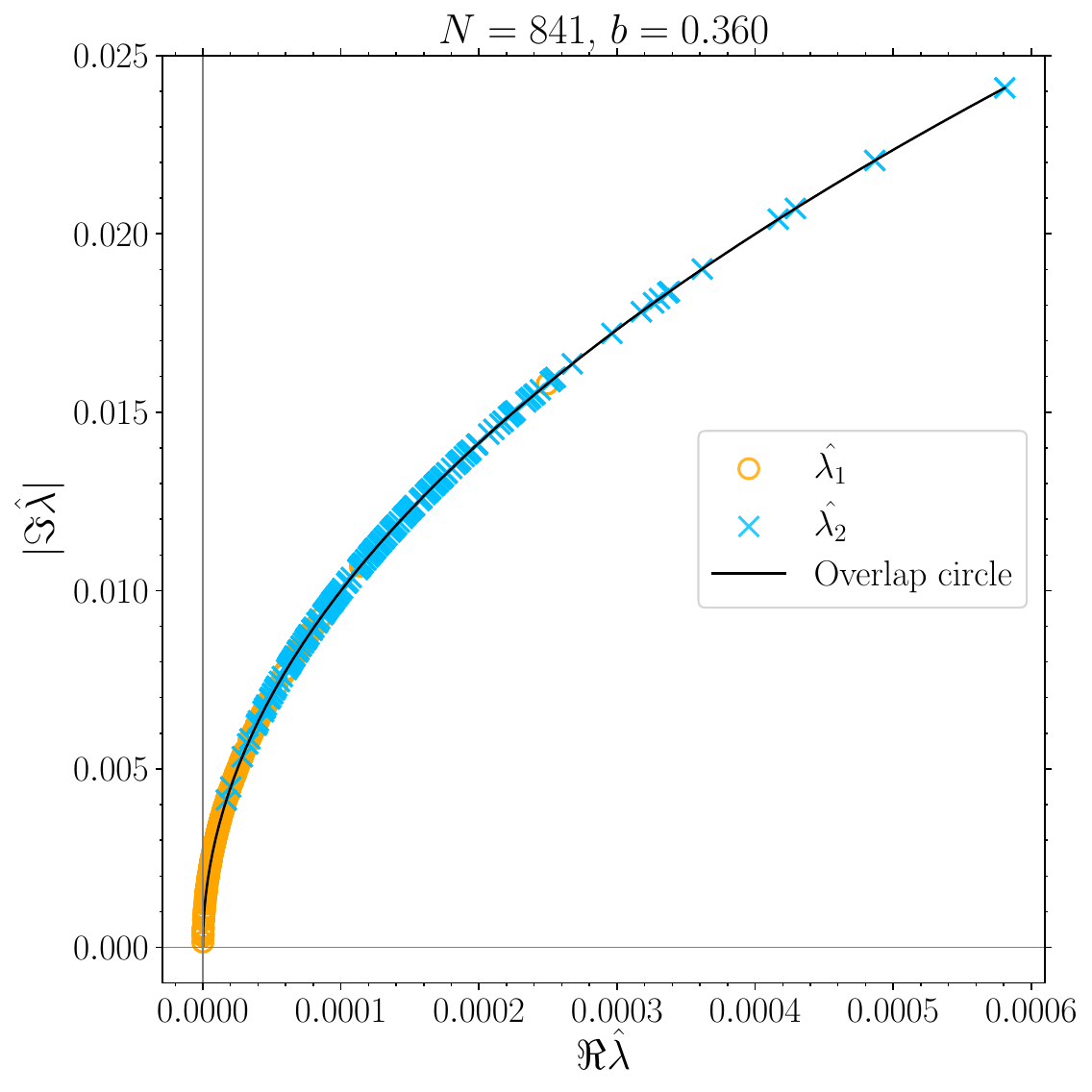}
\caption{First two complex eigenvalues of the truncated overlap Dirac operator obtained with $M=1.2$, $N_5=24$ and stout-smeared gauge links with stout parameters $\rho=0.1$, $n_s=5$, corresponding to a smoothing radius of two lattice spacings. The solid black line stands for the expected overlap circle~\eqref{eq:overlap_circle} in the complex plane. Left panel: $N=529$, $b=0.355$. Right panel: $N=841$, $b=0.360$.}
\label{fig:spectrum}
\end{figure}

Another interesting test of the good chiral properties of our truncated overlap Dirac operator is to check that lattice eigenvalues lie on the overlap circle:
\beq\label{eq:overlap_circle}
\hat{\lambda} = \frac{1}{2} + \frac{1}{2} \ee^{\ii\theta},
\eeq
with $\hat{\lambda}$ the eigenvalue of the massless overlap operator. In Fig.~\ref{fig:spectrum} we plot as en example the first two complex eigenvalues $\hat{\lambda}_1$ and $\hat{\lambda}_2$. Due to $\gamma_5$-hermiticity, eigenvalues come in pairs that have the same real part and opposite imaginary parts. Thanks to this symmetry, we can just limit to plot the upper half of the circle in the complex plane. The plots in Fig.~\ref{fig:spectrum} refer to the cases $N = 529$ at $b=0.355$ and $N=841$ at $b=0.360$, corresponding to similar effective sizes in physical units:
\beq
b = 0.355, \, N=529 &\quad\implies\quad& \ell \sqrt{\sigma} = (a \sqrt{\sigma})\sqrt{N} \simeq 5.54,\\
\nonumber\\[-1em]
b = 0.360, \, N=841 &\quad\implies\quad& \ell \sqrt{\sigma} = (a \sqrt{\sigma})\sqrt{N} \simeq 5.97.
\eeq
As it can be observed, all eigenvalues of all configurations considered indeed lie on the overlap circle for $\Nfive=24$ and our choice of smearing parameters. It is worth commenting that the eigenvalues displayed in the plot were obtained by choosing the free parameter $M$ appearing in the overlap operator as $M=1.2$. After checking for the $N=289$ ensemble that $M=1.5$ gave perfectly compatible results for the eigenvalue distributions, we kept this choice throughout all calculations.

\subsection{Scale-invariant RMT predictions}

To test the agreement between our overlap eigenvalues and Random Matrix Theory (RMT), we analyzed the ratios defined in Eq.~\eqref{eq:ratio_rmt}, and quantified their deviation from unity. In order to compute the expectation values $\braket{\lambda_k}$, we computed the first few low-lying eigenvalues of the massless overlap operator for a few hundred gauge configurations, and we identified the physical Dirac eigenvalue $\lambda$ with the absolute value of the overlap one $\lambda_{\rm o}$:
\beq
\lambda \longrightarrow \vert \lambda_{\rm o} \vert, \qquad \qquad a\lambda_{\rm o} = M(2-M){\hat{\lambda}}.
\eeq
Clearly, we could have also opted for the imaginary part of $\lambda_{\rm o}$ as well, being the real part just a lattice artifact that vanishes in the continuum limit. However, we practically observed no numerical difference between $\braket{\Im\{\lambda_{\rm o}\}}$ and $\braket{\vert \lambda_{\rm o}\vert}$, as $\Re\{\lambda_{\rm o}\}$ turned out to be always at least one order of magnitude smaller than $\Im\{ \lambda_{\rm o}\}$ for all computed eigenvalues, see Fig.~\ref{fig:spectrum}. Concerning the RMT predictions $\langle z_{k_1}\rangle_{\RMT} /  \langle z_{k_2}\rangle_{\RMT}$, the numerical evaluations of the analytic results for these ratios are taken from Tab.~VII of Ref.~\cite{Bernardoni:2008ei}. Finally, concerning the topological sector, we did not find any exact zero-mode in our spectra, thus, we compared our data with RMT predictions in the $Q=0$ topological sector.

\begin{figure}[!t]
\centering
\includegraphics[scale=0.45]{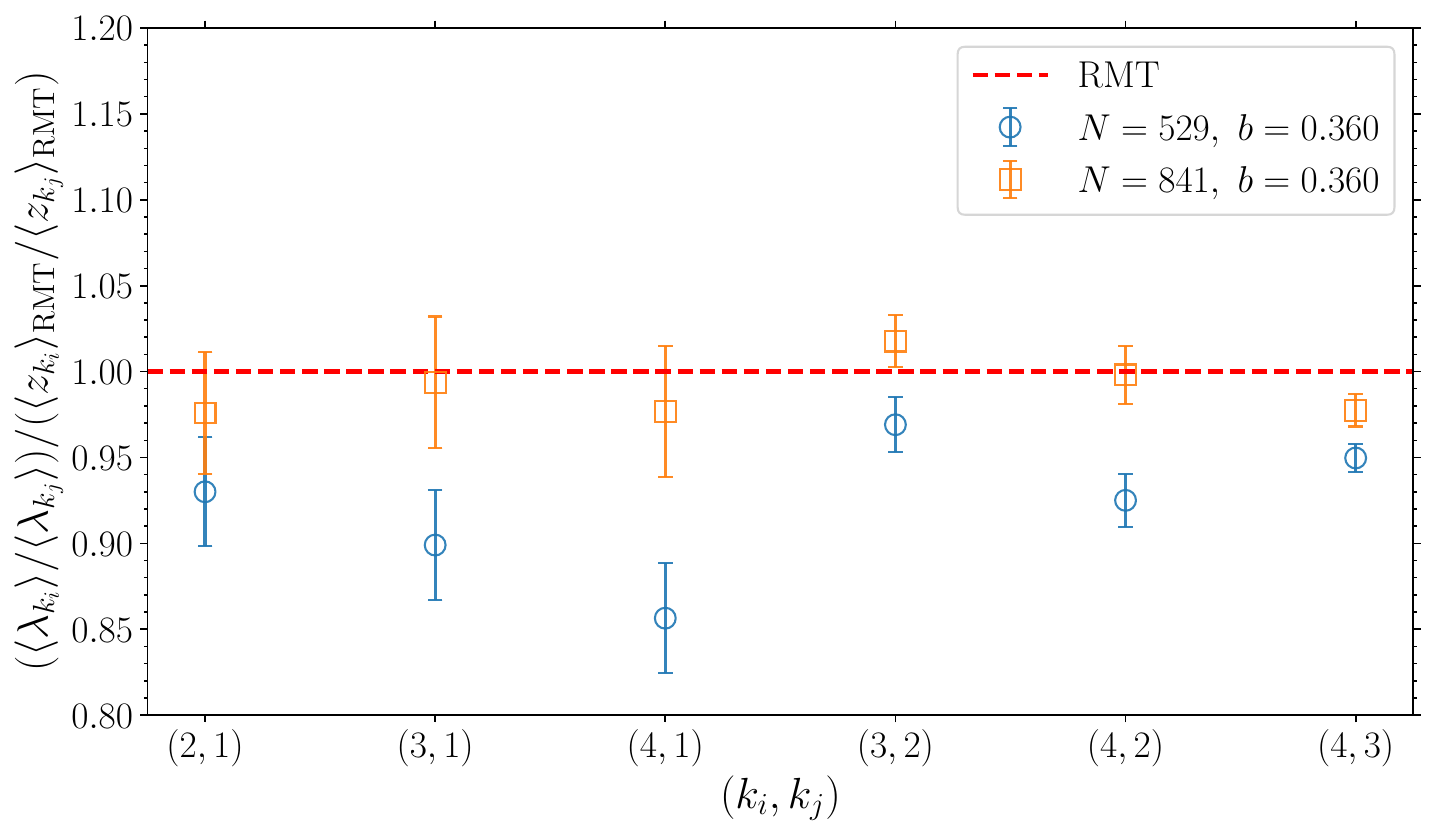}
\includegraphics[scale=0.45]{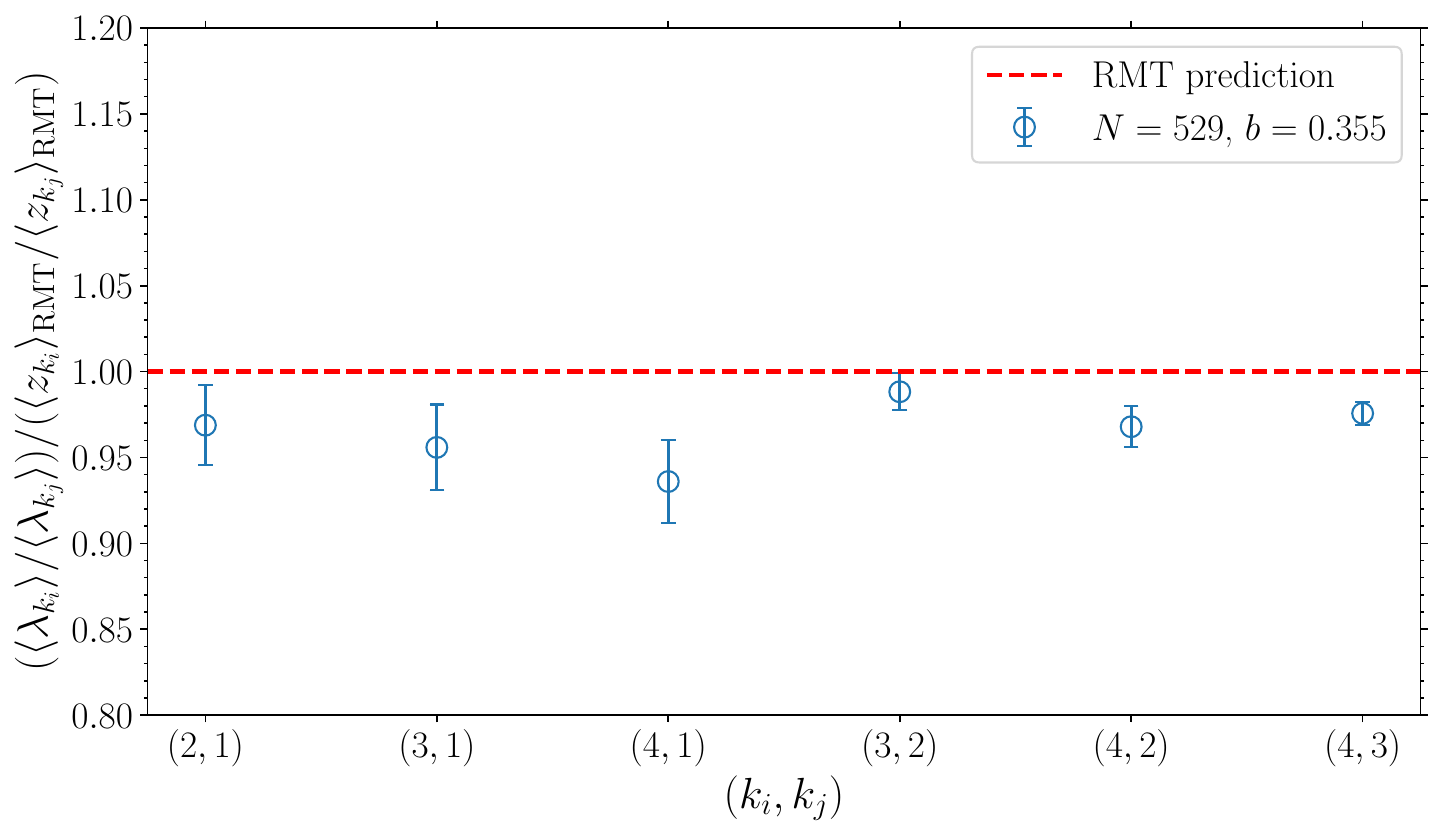}
\caption{Ratios of expectation values $\braket{\lambda_{k_1}}/\braket{\lambda_{k_2}}$, obtained considering all combinations with $k_i,k_j\le 4$, divided by the same quantity computed in the RMT model, $\braket{z_{k_1}}_{\RMT}/\braket{z_{k_2}}_{\RMT}$. The red dashed line at $y=1$ represents the agreement between lattice data and RMT predictions. Top and bottom panels refer, respectively, to $b=0.360$ and $b=0.355$.}
\label{fig:rmt_ratios}
\end{figure} 

\begin{figure}[!t]
\centering
\includegraphics[scale=0.52]{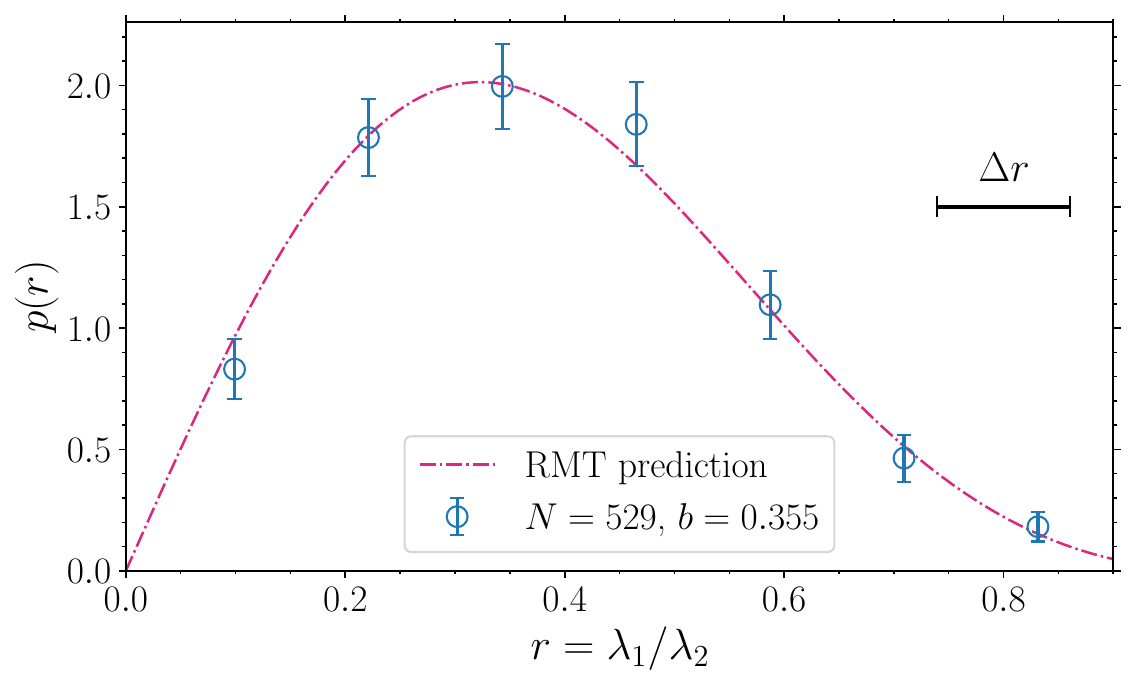}
\includegraphics[scale=0.52]{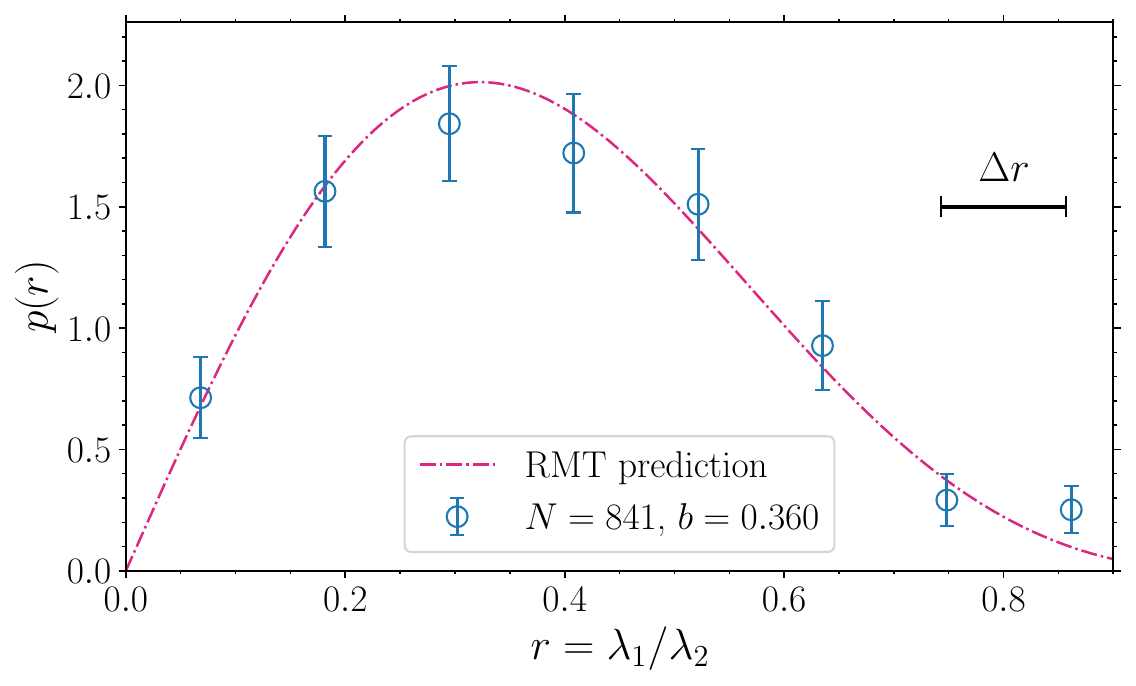}
\caption{Distribution of the ratio of the first two Dirac eigenvalues,
$r=\lambda_{1}/\lambda_{2}$. Top panel: $N=529$, $b=0.355$. Bottom panel: $N=841$, $b=0.360$. The dashed curve represents the parameter-free functional form
predicted by RMT for $p(r)$, reported in Eq.~\eqref{eq:p_r_prediction}. In all plots we also show the constant bin size $\Delta r$ used to obtain the displayed distributions.}
\label{fig:rmt_ratio_distr}
\end{figure}

In Fig.~\ref{fig:rmt_ratios} we plot the results obtained for $b=0.355$ and $b=0.360$, for several choices of $k_1$ and $k_2$. Let us first comment the $b=0.360$ case, where two values of $N$ are reported. As it can be seen, for $N=529$, corresponding to $\ell\sqrt{\sigma}\simeq 4.73$, one can clearly observe deviations from RMT predictions, which can become as large as $\sim 15\%$. Smaller values of $N=289,361$, not shown in the plot, exhibit even larger deviations. On the other hand, when $N$ is increased to $841$, corresponding to the larger size $\ell\sqrt{\sigma}\simeq 5.97$, we observe perfect agreement with RMT predictions within errors in all examined cases, up to $k_1,k_2=4$. For the coarser point $b=0.355$, where $N=529$ corresponds to $\ell\sqrt{\sigma}=5.54$ (i.e., an intermediate volume between the two earlier discussed), we observe small deviations of the order of a few percent ($\sim 5\%$ at most). Overall, thus, the collected evidence shows that finite-volume effects play a crucial role in the present analysis, and that the agreement between lattice data and RMT predictions is controlled by the effective size in physical units. As $\ell \to \infty$, i.e., as $N\to \infty$ at fixed $b$, RMT predictions become more and more accurate.

Analogous conclusions can be drawn by comparing the lattice distribution of the ratio of eigenvalues $r=\lambda_1/\lambda_2$ with the parameter-free RMT prediction for $p(r)$ in Eq.~\eqref{eq:p_r_prediction}. As it can be seen in Fig.~\ref{fig:rmt_ratio_distr}, the numerical results for the normalized distribution of $r$,
\beq
p(r)=\frac{\braket{\text{number of $\frac{\lambda_1}{\lambda_2}$} \in [r-\frac{1}{2}\Delta r,r+\frac{1}{2}\Delta r]}}{n_{\rm conf} \, \Delta r},
\eeq
obtained for the two largest effective sizes explored --- namely, $N=529$ at $b=0.355$ and $N=841$ at $b=0.360$ --- agree very well with expectations. We stress again that the curves shown in Fig.~\ref{fig:rmt_ratio_distr} are scale-invariant RMT predictions with no free parameter, and thus there is no fitting procedure involved in the comparison with numerical data.

\subsection{Parameter-dependent RMT predictions}

Our previous analysis showed that an effective size $\ell\sqrt{\sigma}\gtrsim 5.5$ is sufficient for the first few Dirac eigenvalues to fall into the RMT regime. The goal of this section is to check that their parameter-dependent probability distribution is indeed described by the analytic RMT functional form with the same parameter $\Sigma$.

In the large-$N$ TEK case, the relation between RMT and Dirac eigenvalues in~\eqref{eq:QCD_RMT_matching} is modified as:
\beq\label{eq:RMT_QCD_eigval_matching_largeN_TEK}
z_k = \lambda_k \frac{\Sigma}{N} V, \qquad V=a^4 N^2,
\eeq
where we performed the substitution $\Sigma\rightarrow\Sigma/N$, which is the quantity that possesses a finite large-$N$ limit, and where we have used the fact that in the TEK model the lattice size $\ell=aL$ is replaced by the effective torus size $\ell=aL=a\sqrt{N}$. There are two possible ways to extract $\Sigma/N$ from RMT predictions:
\begin{enumerate}[\textbf{(\arabic*)}]
\item \label{list:method1}
Performing a best fit of the normalized lattice probability distribution of $a\lambda_k$ to the expected RMT shapes assuming the following relation:
\beq\label{eq:eigvals_relation_latunits}
z_k = (a\lambda_k) A, \qquad A = a^3 \frac{\Sigma}{N} N^2,
\eeq
with $A$ the only fit parameter. For example, the normalized RMT probability distribution for the first eigenvalue $x_1\equiv a \lambda_1$ assuming~\eqref{eq:eigvals_relation_latunits} reads:
\beq
p_1(x_1) =  \frac{1}{2}(A^2 x_1)\ee^{-\frac{1}{4} A^2 x^2_1} \ .
\eeq
This strategy, adopted in~\cite{Narayanan:2004cp,Gonzalez-Arroyo:2005dgf}, does not only allow to extract the condensate, but also constitutes a strong test of the agreement between lattice data and RMT predictions.
\item \label{list:method2}
From Eq.~\eqref{eq:RMT_QCD_eigval_matching_largeN_TEK}, computing this ratio of expectation values:
\beq\label{eq:sigma_from_ratio}
a^3 \frac{\Sigma}{N} = \frac{\braket{z_k}_{\RMT}}{N^2\braket{a\lambda_k}},
\eeq
where
\beq
\langle z_{k} \rangle_{\RMT}= \int_0^\infty z_{k} \, p_k(z_{k}) \, \dd z_{k}.
\eeq
This strategy~\ref{list:method2} has been applied in previous standard QCD studies to extract the condensate from RMT predictions~\cite{Giusti:2003gf,Wennekers:2005wa,Bernardoni:2008ei,Hasenfratz:2008ce,Bernardoni:2010nf,Deuzeman:2011dh,Catillo:2017qbz}. Clearly, it should provide compatible estimates for $\Sigma/N$ with respect to~\ref{list:method1}.
\end{enumerate}

\begin{figure}[!t]
\centering
\includegraphics[scale=0.39]{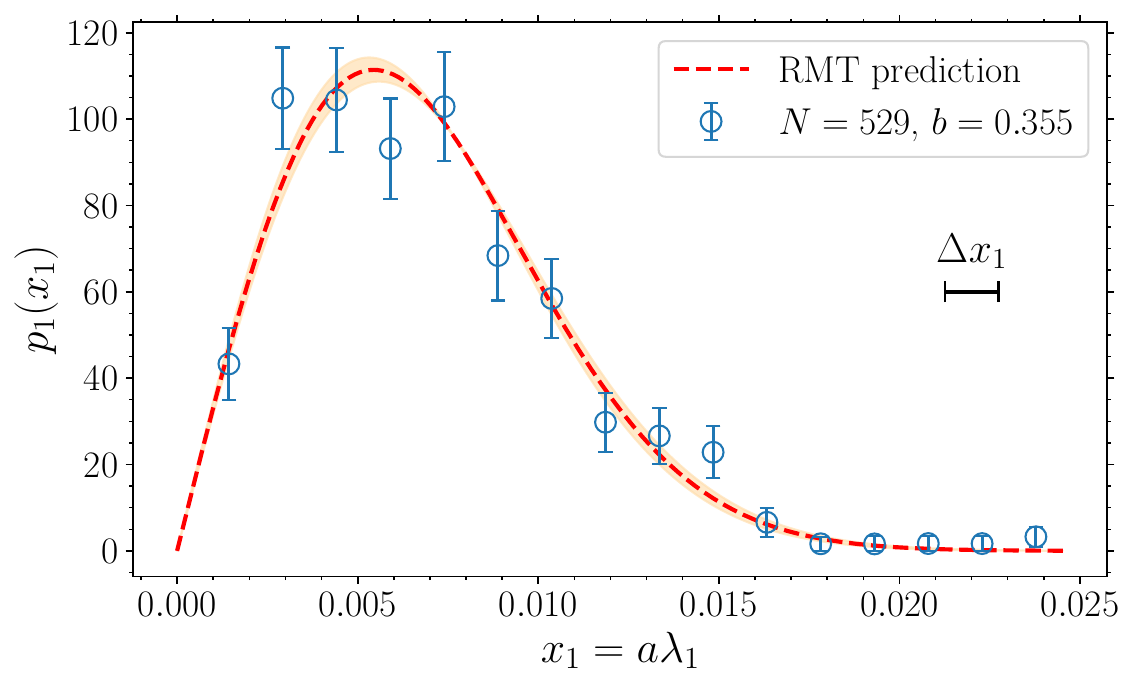}
\includegraphics[scale=0.39]{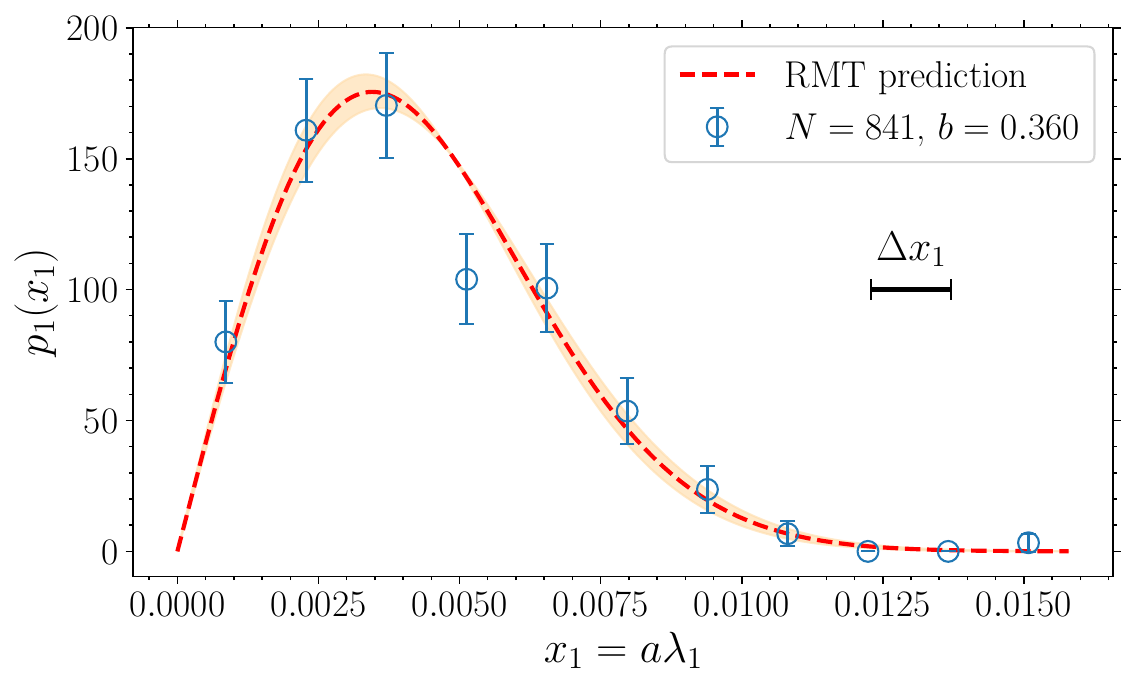}
\includegraphics[scale=0.39]{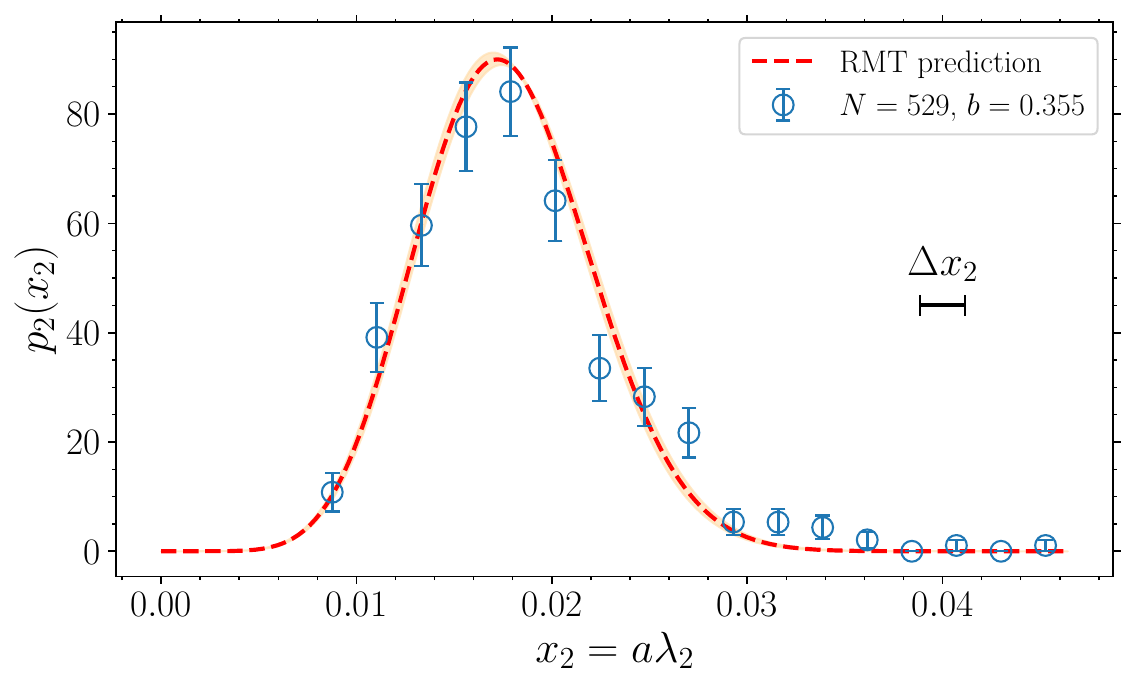}
\includegraphics[scale=0.39]{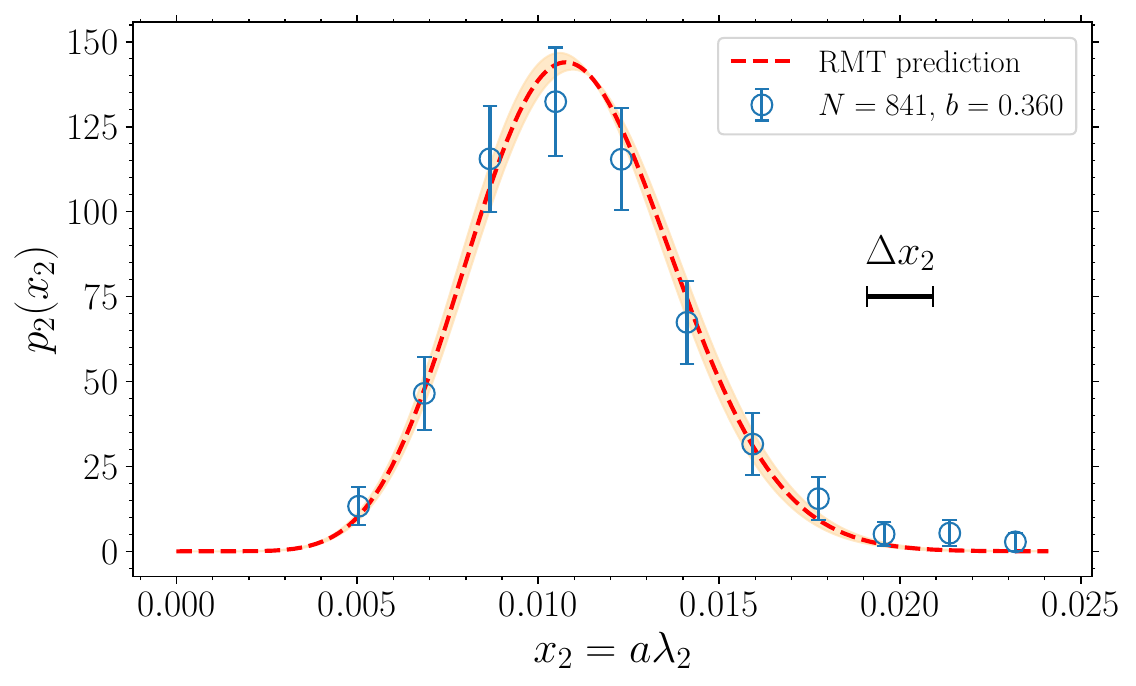}
\caption{Best fits of the probability distribution of the first (top panels) and the second (bottom panels) eigenvalues according to, respectively Eqs.~\eqref{eq:p_1_prediction} and~\eqref{eq:p_2_prediction}. Left panels refer to $N=529,b=0.355$, right panels to $N=841,b=0.360$. In all plots we also show the constant bin sized $\Delta x_1$ and $\Delta x_2$ used to obtain the displayed distributions.}
\label{fig:dist_fit}
\end{figure}

\begin{table}[!t]
\begin{center}
\begin{tabular}{|c|c|c|c|c|c|c|}
\hline
& $N$ & $b$ & $ \tilde\chi^2 $ & $p$-value & \makecell{$10^3\times a^3\Sigma/N$\\$[$RMT fit]} & \makecell{$10^3\times a^3\Sigma/N$\\$[$Eq.~\eqref{eq:sigma_from_ratio}]} \\
\hline
\hline
\multirow{2}{*}{$\lambda_1$}
& 529 & 0.355 & 0.99 & 46\% & 0.891(22) & 0.870(24)\\
& 841 & 0.360 & 0.89 & 52\% & 0.556(20) & 0.561(22)\\
\hline
\hline
\multirow{2}{*}{$\lambda_2$}
& 529 & 0.355 & 1.90 &  2\% & 0.931(12) & 0.899(14)\\
& 841 & 0.360 & 0.55 & 86\% & 0.587(10) & 0.575(11)\\
\hline
\end{tabular}
\end{center}
\caption{Extraction of the large-$N$ condensate from the first two low-lying eigenvalues, both from the best fit to RMT predictions for their probability distribution and via Eq.~\eqref{eq:sigma_from_ratio}.}
\label{tab:res_fit_distr_vs_ratio}
\end{table}

For the two largest effective volumes ($N=529$ for $b=0.355$ and $N=841$ for $b=0.360$), best fits of lattice data for the probability distribution of the first few eigenvalues yield satisfactory reduced chi-squared $\tilde\chi^2\equiv\chi^{2}/n_{\mathrm{dof}}$, confirming the quality of the RMT description. As an example, we show in Fig.~\ref{fig:dist_fit} the best fits of the numerical probability distributions of the first two eigenvalues, $\lambda_1$ and $\lambda_2$. We report the results for the chiral condensate (i.e., the fit parameter $A$) extracted from the best fit in Tab.~\ref{tab:res_fit_distr_vs_ratio}, along with the compatible determinations obtained from the ratio of expectation values in Eq.~\eqref{eq:sigma_from_ratio}. As it can be seen, the estimates obtained from $\lambda_1$ and $\lambda_2$ perfectly agree within errors, in agreement with RMT predictions. On the other hand, when lowering the effective volume, probability distributions start to show deviations from RMT, and the condensates extracted from different choices of $\lambda_k$ are clearly different from each other, cf.~Tab.~\ref{tab:condensate_value_vs_N}.

\begin{table}[!t]
\begin{center}
\begin{tabular}{|c|c|c|c|c|c|}
\hline
$b$ & $N$ & \makecell{$10^3\times a^3\Sigma/N$\\$[\text{From } \lambda_1]$} & \makecell{$10^3\times a^3\Sigma/N$\\$[\text{From } \lambda_2]$} & \makecell{$10^3\times a^3\Sigma/N$\\$[\text{From } \lambda_3]$} & \makecell{$10^3\times a^3\Sigma/N$\\$[\text{From } \lambda_4]$} \\
\hline
\multirow{4}{*}{0.360} & 289 & 0.516(15) & 0.672(10) & 0.7823(69) & 0.8940(64) \\
& 361 & 0.510(22) & 0.605(13) & 0.6681(96) & 0.7550(88) \\
& 529 & 0.537(21) & 0.580(13) & 0.598(11) & 0.6268(83) \\
& 841 & 0.561(22) & 0.575(11) & 0.565(11) & 0.5764(81) \\
\hline
\end{tabular}
\end{center}
\caption{Bare chiral condensate in lattice units obtained from Eq.~\eqref{eq:sigma_from_ratio} for several choices of $k$, and as a function of $N$.}
\label{tab:condensate_value_vs_N}
\end{table}

The differences observed at smaller values of $N$ in the condensates extracted from higher eigenvalues, as discussed in the previous section, can be regarded as finite-volume effects due to not being yet in the regime where RMT provides a reliable description. It is known that RMT is expected to hold in the so-called $\epsilon$-regime of QCD~\cite{Gasser:1987ah}, see, e.g., Refs.~\cite{Verbaarschot:2000dy,Basile:2007ki}. Customarily, QCD is studied in the so-called $p$-regime, obtained taking first the thermodynamic limit $L\to\infty$ at fixed and finite quark mass and then the chiral limit, so that $m_\pi L\to\infty$. In the $\varepsilon$-regime, instead, $m\to0$ and $L\to\infty$ at fixed $m\Sigma V\lesssim\mathcal{O}(1)$. Since $m\Sigma V = 2(F_\pi m_\pi L^2)^2$, this means that $m_\pi L \to 0$ while $F_\pi L\to\infty$. It was shown in Ref.~\cite{Gasser:1987ah} that finite-volume corrections to the thermodynamic limit in the $\epsilon$-regime are power-like, and can be expanded in series of the variable $1/(F_\pi^2 L^2) = 1/(F_\pi^2 \sqrt{V})$. This behavior is very different from the customary exponential suppression of finite-size effects in the $p$-regime, and is related to the peculiar way of taking the chiral and thermodynamic limits.

\begin{figure}[!t]
\centering
\includegraphics[scale=0.3]{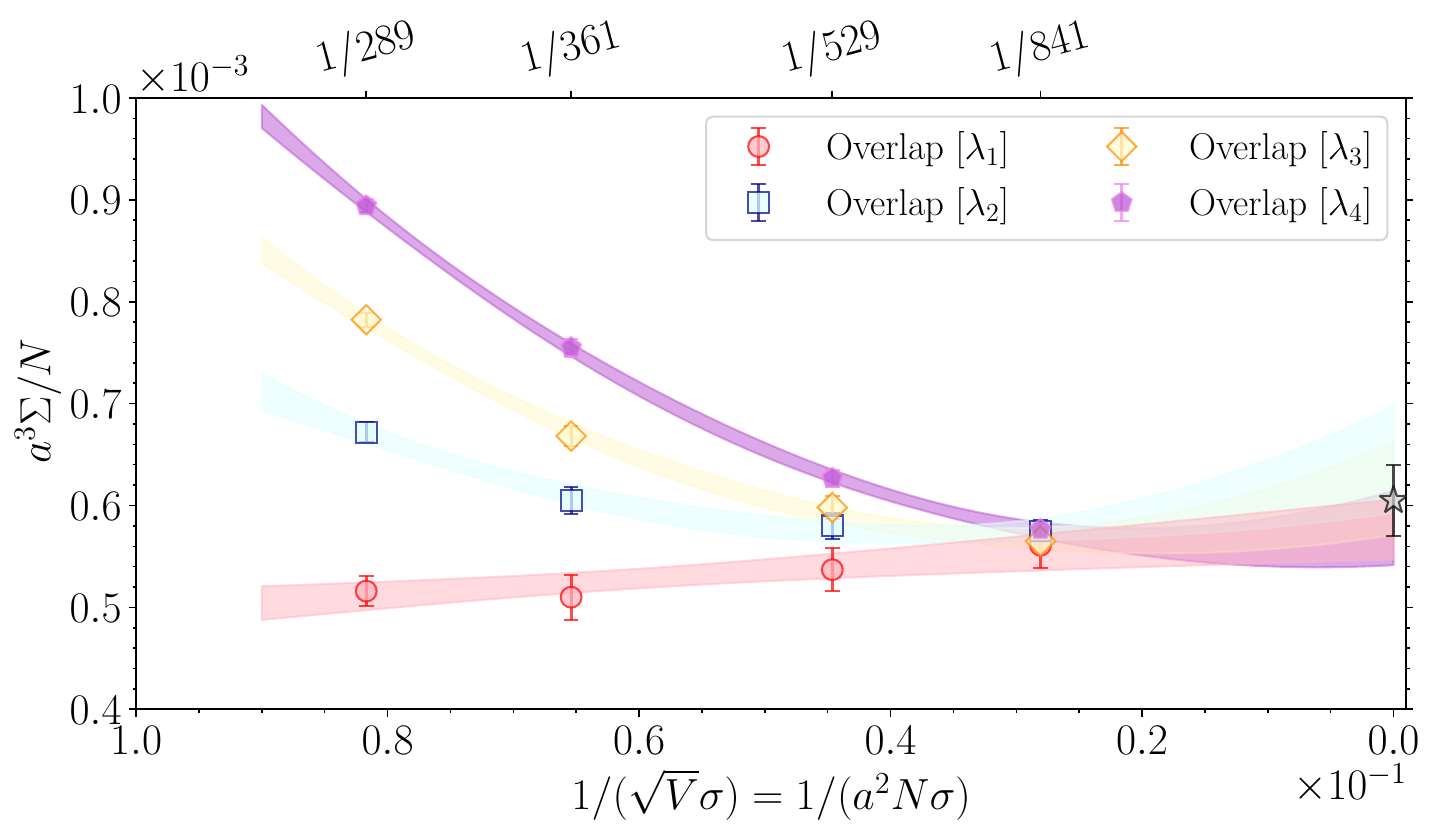}
\includegraphics[scale=0.3]{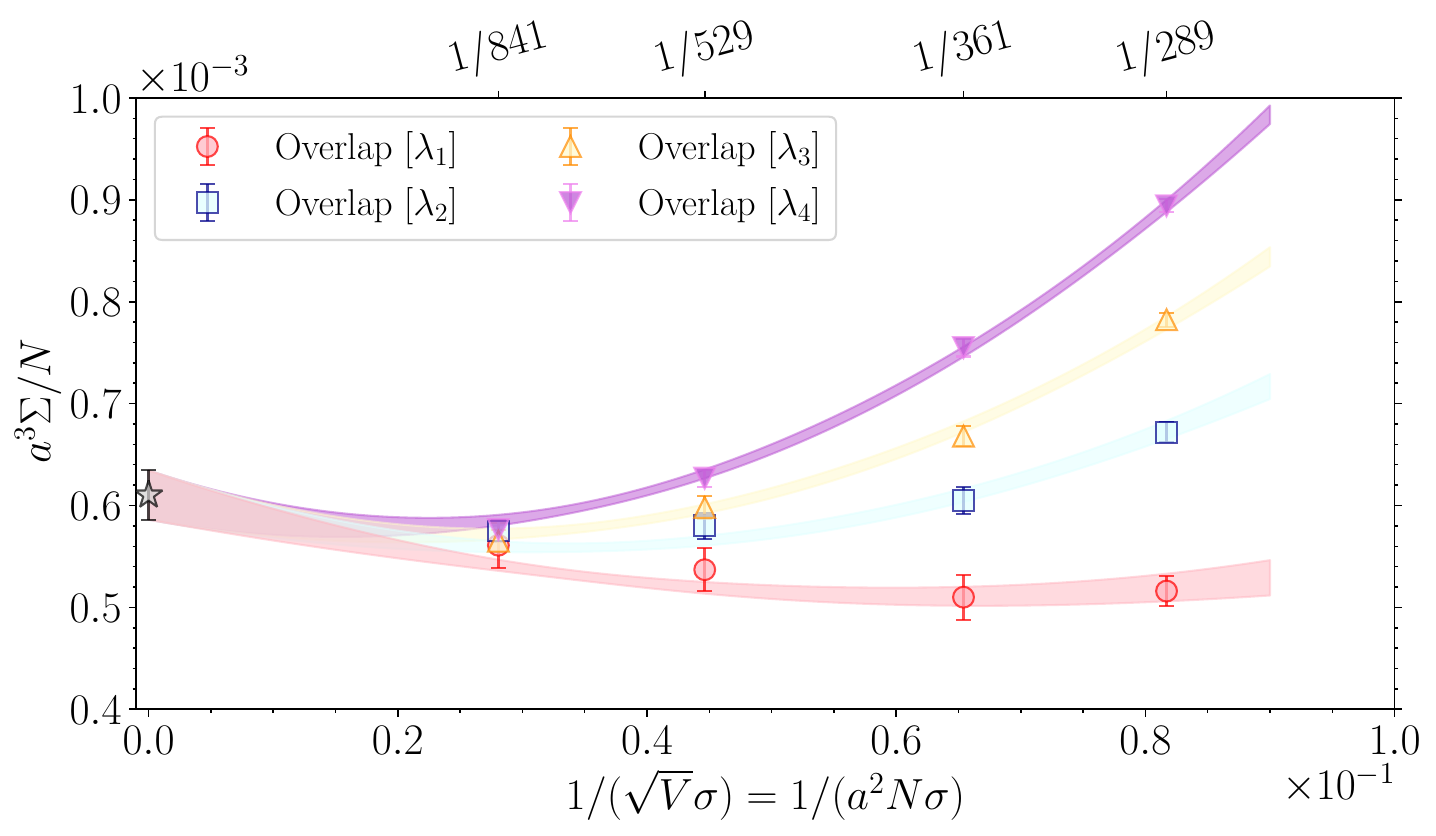}
\caption{The bare large-$N$ chiral condensate $\Sigma/N$ in lattice units as a function of $1/\sqrt{V}$. Left panel: shown curves are polynomial fits in $1/\sqrt{V}$ to each individual data set. Right panel: shown curves are the result of a global best fit according to the empirical ansatz in Eq.~\eqref{eq:bestfit_cc_vs_N} (bottom panel).}
\label{fig:condensate_vs_N}
\end{figure}

We empirically tried to check whether the volume-dependence of our data can be described by polynomials in the variable $x=1/\sqrt{V}$ or not, with $V=a^2N^2$ the effective volume. Interestingly, our results for the condensate for $b=0.360$ as a function of $N$ obtained from the first four low-lying eigenvalues can be all well described by polynomial curves in this variable, extrapolating to compatible values at $x=0$, and with similar slopes close to zero, see Fig.~\ref{fig:condensate_vs_N} (left panel). From these fits we can quote the following final result:
\beq\label{eq:cc_final}
\frac{a^3 \Sigma}{N} = 0.605(35)\times 10^{-3}.
\eeq
where the central value is the average of the four extrapolations, and the quoted error is the maximum semi-dispersion observed among them. The observed behavior of the best fits performed on a $k$-by-$k$ basis also lead us to try a global best fit of the data, assuming the following functional form:
\beq\label{eq:bestfit_cc_vs_N}
\frac{a^3\Sigma_k}{N} = B_0 \left(1 + B_1 \frac{x}{\sigma} + B_2^{(k)} \frac{x^2}{\sigma^2}\right), \qquad x = \frac{1}{\sqrt{V}} = \frac{1}{a^2N},
\eeq
where $\Sigma_k$ is the condensate extracted from $\lambda_k$, and where we impose a common large-volume limit $B_0$ and a common slope $B_1$. This functional form describes our data very well, giving $\tilde{\chi}^2\simeq 0.83$, see also Fig.~\ref{fig:condensate_vs_N} (right panel). Moreover, it gives a compatible thermodynamic limit with respect to the one obtained from the unconstrained fit:
\beq\label{eq:cc_final_other1}
\frac{a^3 \Sigma}{N} = 0.610(24)\times 10^{-3}.
\eeq
Although both extrapolated results are larger, they are nevertheless compatible within errors with the average of the $N=841$ results:
\beq\label{eq:cc_final_other2}
\frac{a^3 \Sigma}{N} = 0.575(25)\times 10^{-3}.
\eeq
where again the error is the sum of a statistical and a systematic one, taking into account the dispersion among the data. In order to be conservative, we will consider the result in Eq.~\eqref{eq:cc_final} as our final overlap determination for $b=0.360$, as this has the largest error among the results in Eqs.~\eqref{eq:cc_final},~\eqref{eq:cc_final_other1} and~\eqref{eq:cc_final_other2}.

Clearly, apart from polynomials in $1/\sqrt{V}$, also other functional forms could equally well describe our data. Thus, to really pinpoint the exact form of finite-size corrections, more theoretical studies would be needed to work out the expected $\epsilon$-regime volume-dependence of QCD low-energy constants in the TEK model. This is left for future studies.

\subsection{Renormalization of the overlap condensate and comparison with Wilson quarks}\label{sec:ZS}

In order to compare our overlap determination of $\Sigma/N$ with the previous TEK one obtained in Ref.~\cite{Bonanno:2025hzr} with Wilson fermions, it is necessary to renormalize $\Sigma$. The overlap valence quark mass and the overlap chiral condensate renormalize as follows:
\beq
\Sigma_{\R} &=& \ZS \Sigma,\\
m_{\R} &=& \Zm m = \frac{1}{\ZS} m,
\eeq
where $am = M(2-M)\hat{m}$, and where we used the relation $\Zm \ZS = 1$, holding by virtue of the lattice chiral symmetry. These renormalization constants depend on the renormalization scheme s and on the renormalization scale $\mu$. From now on, any presented renormalized quantity will be expressed according to the standard choices of renormalization scheme and scale: $\mathrm{s}=\overline{\mathrm{MS}}$ and $\mu = 2$ GeV. In this work we assume $\mu=2$ GeV $\longrightarrow \mu/\sqrt{\sigma}=3.75$ as in the previous large-$N$ studies~\cite{DeGrand:2023hzz,Bonanno:2025hzr}.

In order to compute $\ZS$, we follow the strategy put forward in Ref.~\cite{Wennekers:2005wa}. In this study, the authors obtained $\ZS$ for overlap quarks from the ratio of the bare overlap mass and the continuum renormalized mass matched at equal values of the pion mass, for a certain reference value $m_\pi=m_\pi^{(\rm ref)}$:
\beq\label{eq:ZS_overlap}
\ZS = \frac{m\,\big\vert_{m_\pi\,=\,m_\pi^{(\rm ref)}}}{m_{\R}\,\big\vert_{m_\pi\,=\,m_\pi^{(\rm ref)}}}.
\eeq
Following again the lines of Ref.~\cite{Wennekers:2005wa}, we compute the continuum renormalized quark mass $m_{\R}$ as a function of $m_\pi$ using Wilson fermion results. Within the TEK model, we computed the Wilson pion mass as a function of the bare Wilson quark mass~\cite{Bonanno:2025hzr}. Concerning non-perturbative determinations of $\ZS$ needed to renormalize the bare Wilson quark mass, in absence of direct TEK determinations, we relied on the finite-$N$ determinations of~\cite{Castagnini:2015ejr}, suitably extrapolated towards $N=\infty$ and interpolated at our values of the lattice spacing. We refer the reader to Ref.~\cite{Bonanno:2025hzr} for more details on this point.

\begin{figure}[!t]
\centering
\includegraphics[scale=0.55]{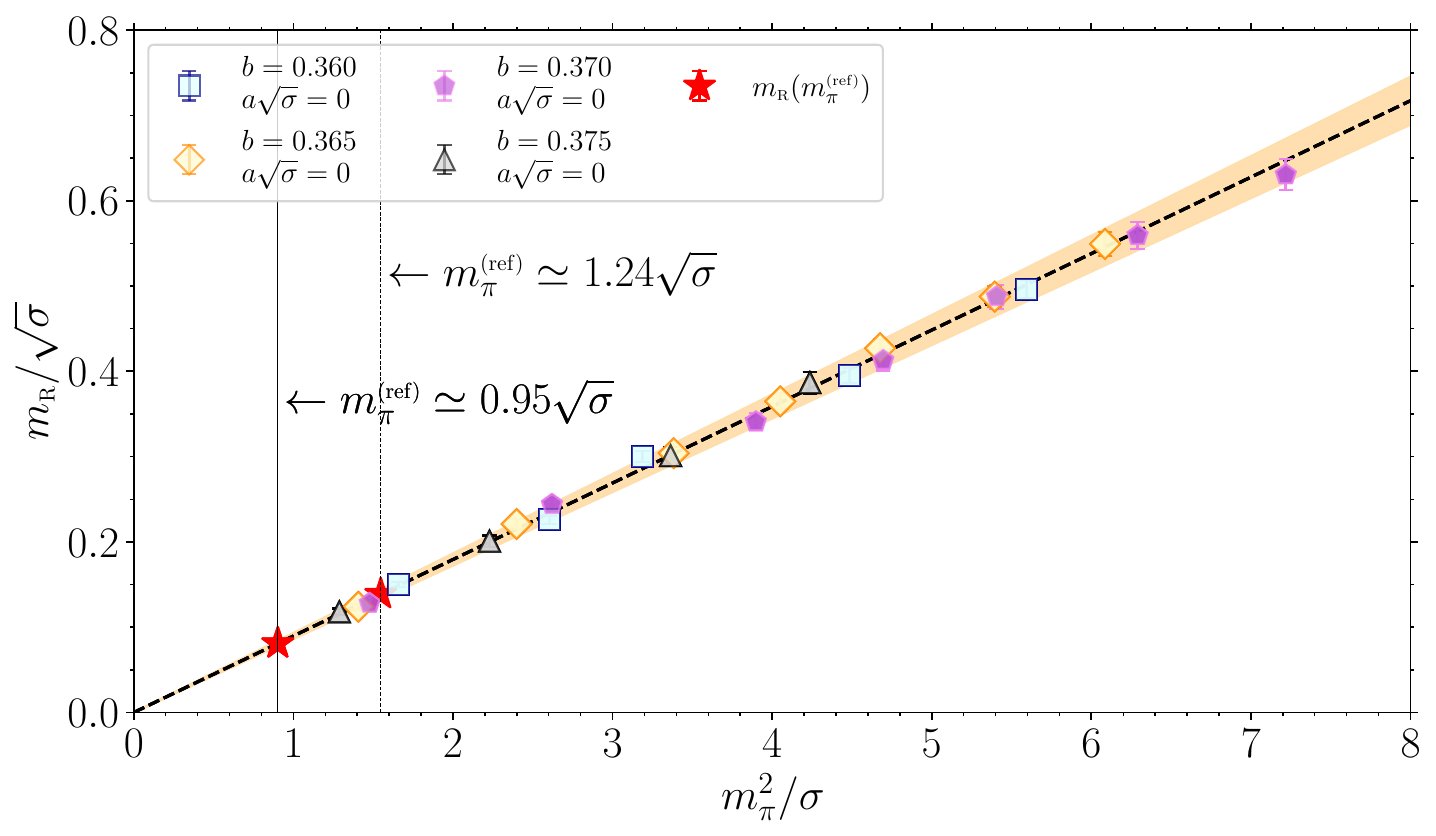}
\caption{Continuum determination of the renormalized quark mass as a function of the pion mass. Fitted data come from~\cite{Bonanno:2025hzr} and have been obtained with Wilson fermions.}
\label{fig:mR_vs_mpi}
\end{figure}

First, we computed the renormalized mass in physical units for several values of the hopping parameter $\kappa$ and of the coupling $b$ (here $\kappa_c$ is the critical hopping parameter):
\beq
\frac{1}{\sqrt{\sigma}}m_\R(b,\kappa) = \frac{1}{\ZS(b)} \frac{1}{a(b)\sqrt{\sigma}}\left(\frac{1}{2\kappa}-\frac{1}{2\kappa_{\rm c}(b)}\right).
\eeq
Then, we performed the following global best fit as a function of $b$ and $\kappa$:
\beq
\frac{1}{\sqrt{\sigma}}m_\R(b,\kappa) = A_1 \, x_1(b) + A_2 \, x_2(b,\kappa) + A_3 \, x_1(b) x_2(b,\kappa),
\eeq
where
\beq
x_1 = a\sqrt{\sigma}, \qquad \quad x_2 = \frac{m_\pi^2}{\sigma},
\eeq
and $A_1$, $A_2$, $A_3$ are fit parameters. Once $A_1$, $A_2$ and $A_3$ are determined, the continuum renormalized quark mass as a function of the pion mass is given by the curve:
\beq\label{eq:mR_vs_mpi_curve}
\frac{1}{\sqrt{\sigma}} m_\R(m_\pi) = A_2 \frac{m_\pi^2}{\sigma}.
\eeq

In Fig.~\ref{fig:mR_vs_mpi} we plot the curve in Eq.~\eqref{eq:mR_vs_mpi_curve}, and on top of it we also show the fitted points ($b=0.360,0.365,0.370,0.375$) to which we have subtracted lattice artifacts (i.e. the terms proportional to $A_1$ and $A_3$). This figure shows the very good quality of the best fit, which has $\tilde{\chi}^2\simeq 0.84$. On a side note, we observe that $1/(2A_2) = B_\R\sqrt{\sigma}$, with $B_\R\sqrt{\sigma}=\Sigma_\R/(F_\pi^2\sqrt{\sigma})$. We find $B_\R/\sqrt{\sigma}=5.58(22)$, in perfect agreement with the determination $B_\R/\sqrt{\sigma}=5.58(26)$ of~\cite{Bonanno:2025hzr} from individual $b$-by-$b$ chiral fits to the quark-mass dependence of the pion mass.

\begin{figure}[!t]
\centering
\includegraphics[scale=1.05]{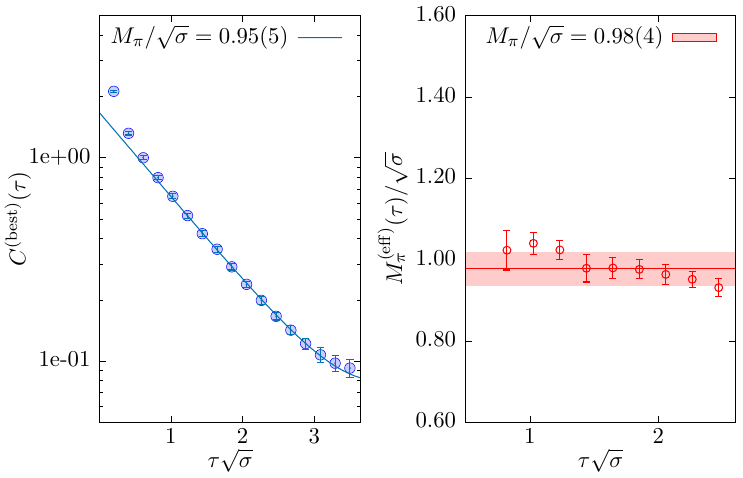}
\caption{Extraction of the pion mass from the exponential decays of the $\pi$ optimal correlator obtained from the resolution of the GEVP (left panel). We also show the plateau in the effective masses (right panel). Plots refer to $N = 361$, $b = 0.360$, $\hat{m}=0.02$, corresponding to $m_\pi \ell = (am_\pi)\sqrt{N} \simeq 3.71$ and $\ell\sqrt{\sigma} = (a\sqrt{\sigma})\sqrt{N}\simeq3.91$.}
\label{fig:pioncorr}
\end{figure}

Finally, we measured the overlap pion mass for $\hat{m}=0.02$, which was found to be:
\beq\label{eq:ref_pion_mass_overlap}
\frac{m_\pi^{\rm (ref)}}{\sqrt{\sigma}} = 0.95(5), \qquad (\hat{m}=0.02 \, \implies \, am=0.0192).
\eeq
This will be our reference matching point to compute $\ZS$ from Eq.~\eqref{eq:ZS_overlap}, and is reported in Fig.~\ref{fig:mR_vs_mpi} as a solid line. This value is obtained from the customary exponential best fit to the pion correlator, cf.~Fig.~\ref{fig:pioncorr} (left panel). This correlation function is obtained from a standard GEVP analysis using an extended basis of smeared pseudo-scalar operators. As a standard cross-check, we verified that the effective mass, obtained by solving the GEVP at all times, exhibits a plateau in the same range where we performed the exponential best fit. A constant fit to this plateau gives the perfectly compatible result $\frac{m_\pi^{\rm (ref)}}{\sqrt{\sigma}} = 0.98(4)$, cf.~Fig.~\ref{fig:pioncorr} (right panel). For more details on the GEVP analysis we performed, we refer the reader to Ref.~\cite{Bonanno:2025hzr}, a dedicated paper on the extraction of the low-lying meson spectrum of large-$N$ QCD from the TEK model. Note also that, since in Ref.~\cite{Bonanno:2025hzr} we did not observe any finite-volume effect in meson masses obtained for $m_\pi\ell\gtrsim 4$ within our percent precision, we expect that any possible $1/V$ finite-volume effect due to the fixed $Q=0$ topology~\cite{Brower:2003yx,Aoki:2007ka} of our gauge configurations is below our statistical accuracy.

The renormalization constant from Eq.~\eqref{eq:ZS_overlap}, matched at the overlap pion and quark mass reported in~\eqref{eq:ref_pion_mass_overlap}, reads:
\beq\label{eq:ZS_overlap_final}
\frac{m_\R}{\sqrt{\sigma}}\bigg\vert_{m_\pi \, =\, m_\pi^{\rm (ref)}} = 0.0809(32) \qquad \implies \qquad \ZS = \frac{am/(a\sqrt{\sigma})}{m_\R/\sqrt{\sigma}} = 1.153(45).
\eeq
In order to check the magnitude of possible finite-size effects affecting our pion mass calculation, we also repeated the matching procedure to obtain $\ZS$ for a heavier pion. In particular, we chose $\hat{m}=0.034$ ($am=0.03264$), and found that it corresponded to $m_\pi/\sqrt{\sigma}=1.244(24)$, i.e., $m_\pi\ell\simeq 4.86$ (which is larger than the smallest value of $m_\pi\ell$ used with Wilson fermions in Ref.~\cite{Bonanno:2025hzr}). Adopting this determination as the reference pion mass leads to $m_\R/\sqrt{\sigma}=0.1388(54)$, cf.~Fig.~\ref{fig:mR_vs_mpi} (dashed line), and eventually to $\ZS=1.143(45)$, which is perfectly compatible with the estimate in Eq.~\eqref{eq:ZS_overlap_final}.

Considering our final result for the bare overlap condensate in lattice units obtained in the previous section,
\beq
\frac{a^3 \Sigma}{N} = 0.605(35) \times 10^{-3},
\eeq
and using $\ZS$ in Eq.~\eqref{eq:ZS_overlap_final}, we finally find, for the renormalized condensate in physical units:
\beq
\frac{\Sigma_\R}{N\sqrt{\sigma^3}} = 0.0800(63) \qquad \text{($b=0.360$, overlap)}.
\eeq
It is now interesting to compare our overlap result with the large-$N$ TEK Wilson determination of~\cite{Bonanno:2025hzr}. On general grounds, overlap fermions are expected to be affected by $\mathcal{O}(a^2)$ leading lattice artifacts, as opposed to Wilson fermions, which instead suffer for $\mathcal{O}(a)$ corrections to the continuum limit. This means that overlap fermions are expected to exhibit a faster convergence towards the continuum limit. This different behavior is due to the fact that they enjoy the lattice chiral symmetry, unlike Wilson fermions. Because of this, overlap and Wilson results obtained at the same value of $b$ do not need to agree. For the same lattice spacing, $b=0.360$, we found with Wilson fermions in the chiral limit~\cite{Bonanno:2025hzr}:
\beq
\frac{\Sigma_\R}{N\sqrt{\sigma^3}} = 0.0711(46) \qquad \text{($b=0.360$, Wilson)}.
\eeq
In the continuum limit, instead:
\beq
\frac{\Sigma_\R}{N\sqrt{\sigma^3}} = 0.0889(23) \qquad \text{(continuum limit, Wilson)}.
\eeq
Our overlap result is indeed much closer to the continuum value with respect to the Wilson one obtained at the same lattice spacing. Our overlap result is thus perfectly consistent with the general theoretical expectation that overlap quarks exhibit a faster convergence towards the continuum limit by virtue of the lattice chiral symmetry.

\section{Conclusions}\label{sec:conclu}

We have presented the first investigation of the universal features of chiral symmetry breaking in large-$N$ QCD from the TEK model, whose employment allowed us to reach $N$ as large as $N=841$.

Adopting a chiral formulation of the Dirac operator, first presented in this study for the TEK model, we showed clear evidence that the low-lying massless Dirac spectrum of large-$N$ QCD follows RMT predictions when the effective size in physical units $\ell=a\sqrt{N}$ is sufficiently large, both by comparing scale-invariant analytic results with numerical data, and by best fitting lattice eigenvalue probability distributions to RMT functional forms. Chiral condensate determinations extracted from the first few low-lying eigenvalues differ at finite $N$, but come together as $N$ is increased, confirming that eventually only one single free-parameter describes them all in the thermodynamic limit, as expected from RMT. After renormalization, our overlap determination of the chiral condensate, despite being obtained for a single value of the lattice spacing, turns out to be very close to the continuum TEK determination obtained in~\cite{Bonanno:2025hzr} from Wilson fermions. This fact is in perfect agreement with the general theoretical expectation that overlap quarks are affected by smaller $\mathcal{O}(a^2)$ lattice artifacts --- compared to the $\mathcal{O}(a)$ affecting Wilson quarks --- by virtue of the lattice chiral symmetry. Clearly, overlap and Wilson determinations will only fully coincide in the continuum limit.

There are many future research directions that could be explored with overlap fermions. It would be very interesting to extend the present study by investigating the quark-mass behavior of the pion mass, as well as of other mesons, or to study the continuum limit of the chiral condensate. The latter would require an ambitious numerical effort, but also a better theoretical characterization of finite-volume effects in the $\epsilon$-regime of the TEK formulation of large-$N$ QCD. Finally, it would be very interesting to extend the use of the chiral lattice Dirac operator to other large-$N$ gauge theories where chiral symmetry breaking plays a crucial role, such as $\mathcal{N}=1$ SUSY Yang--Mills, where the calculation of the gluino condensate is of great theoretical interest.

\section*{Acknowledgements}
It is a pleasure to thank Carlos Pena for many fruitful discussions and for reading this manuscript. We also acknowledge useful discussions with Pilar Hern\'andez and Rajamani Narayanan. This work is partially supported by the Spanish Research Agency (Agencia Estatal de Investigaci\'on) through the grant IFT Centro de Excelencia Severo Ochoa CEX2020-001007-S and, partially, by the grants No. PID2021-127526NB-I00 and PID2024-160152NB-I00, both funded by MCIN/AEI/10.13039/ 501100011033. It is also partially funded by the European Commission – NextGenerationEU, through Momentum CSIC Programme: Develop Your Digital Talent. K.-I.~I.~is supported in part by MEXT as ``Feasibility studies for the next-generation computing infrastructure''. This research was supported in part by grant NSF PHY-2309135 to the Kavli Institute for Theoretical Physics (KITP). We acknowledge support from the SFT Scientific Initiative of INFN. This work was partially supported by the Simons Foundation grant 994300 (Simons Collaboration on Confinement and QCD Strings). Numerical calculations have been performed on the \texttt{Finisterrae~III} cluster at CESGA (Centro de Supercomputaci\'on de Galicia), on the \texttt{Drago} cluster at CSIC (Consejo Superior de Investigaciones Científicas) and on the Hydra and Ciclope clusters at IFT. We acknowledge HPC support by Emilio Ambite, staff hired under the Generation D initiative, promoted by Red.es, an organization attached to the Ministry for Digital Transformation and the Civil Service, for the attraction and retention of talent through grants and training contracts, financed by the Recovery, Transformation and Resilience Plan through the European Union's Next Generation funds. We have also used computational resource of Oakbridge-CX, at the University of Tokyo through the HPCI System Research Project (Project ID: hp230021 and hp220011), of Cygnus at Center for Computational Sciences, University of Tsukuba, of SQUID at D3 Center of Osaka university through the RCNP joint use program.

\appendix
\section*{Appendix}
\section{The spectrum of the free overlap operator}
In this Appendix we discus how to restore physical units in the lattice overlap operator $D$, leading to the rescaling formulas presented in Eqs.~\eqref{eq:M_rescaling_phys_units}--\eqref{eq:scale}.

Let us start by generically parameterizing the free massless overlap operator in Fourier space by:
\beq
D_0(p) = \frac{1 + V^{\rm (free)}(p)}{2} = \frac{1}{2} + \frac{B(p) + \ii \sum_\mu \gamma_\mu A_\mu(p) }{2 \left [B^2(p) + \sum_\mu A_\mu^2(p)\right]^{1/2}}\, .
\eeq
From this formula one can easily check that the eigenvalues of the operator $H^2= D_0^\dagger D_0$, with $H=\gamma_5 D_0$, are given by:
\beq
\tilde \lambda^2 = \frac{1}{2} + \frac{ B(p) }{2\left[B^2 + \sum_\mu A_\mu^2(p)\right]^{1/2}}.
\eeq
Moreover, making use of the Ginsparg--Wilson relation:
\beq
D_0 \gamma_5 + \gamma_5 D_0 = 2 D_0 \gamma_5 D_0,
\eeq
and the fact that $\gamma_5 D_0 \gamma_5 = D_0^\dagger$, one can relate the eigenvalues of $D_0$ with $\tilde \lambda^2$ using that,
\beq
D_0  + D_0^\dagger =  2 H^2,
\eeq
therefore $\Re \{ \hat \lambda\} = \tilde \lambda^2$. Making use of the fact that the eigenvalues of $D_0$ lie on a circle and writing
\beq
\hat \lambda= \frac{1}{2} (1 + \ee^{\ii\theta})\, ,
\eeq
it is trivial to derive the form of the imaginary part of the eigenvalues, leading to:
\beq
\hat \lambda= \tilde \lambda^2 \pm \ii \tilde \lambda \sqrt{1-\tilde \lambda^2}\, .
\eeq

\noindent One can now specialize to the case of the overlap operator with kernel given by~\eqref{eq:kernel}. For this, one should first note that the spectrum of the free Wilson--Dirac operator $\gamma_5 \DW(-M)$ in momentum space in the TEK formulation is identical to the standard one, defined on a lattice of $(\sqrt{N})^4$ sites. Using this, one can easily derive the expressions corresponding to $A_\mu(p)$ and $B(p)$ for the overlap kernel used in our work:
\beq
A_\mu (p) &=& \frac{2 s_\mu} {s^2 + (2 + 2 \hat s^2 -M)^2},\\
\nonumber\\
B(p) &=& \frac{s^2 + (2 \hat s^2 -M) (2 + 2 \hat s^2 -M)}{s^2 + (2 + 2 \hat s^2 -M)^2},
\eeq
where $s_\mu = \sin(a p_\mu)$, $s^2 = \sum_\mu \sin^2 (a p_\mu)$ and $\hat s^2 = \sum_\mu \sin^2 (a p_\mu/2)$, with $ap_\mu = 2 \pi n_\mu /\sqrt{N}$, $n_\mu = 0, \cdots, \sqrt{N}-1$. These results agree with those quoted for the standard overlap operator with identical kernel in Ref.~\cite{Aoki:2001su}.

Once we know the expressions for the eigenvalues, the scaling factor in Eq.~\eqref{eq:scale} is obtained by taking into account, for instance, that the eigenvalue corresponding to $H^2$ in the continuum should be $|p|^2$. The overall scaling factor is easily determined by taking the limit of small $a p_\mu$ of $\tilde \lambda$, which leads to: 
\beq
\tilde \lambda^2 \underset{a \,\rightarrow \,0}{\longrightarrow}\, \frac{a^2 |p|^2 }{M^2(2-M)^2}\, ,
\eeq
i.e., the factor of $M(2-M)/a$ appearing in Eqs.~\eqref{eq:M_rescaling_phys_units}--\eqref{eq:scale}.

\providecommand{\href}[2]{#2}\begingroup\raggedright\endgroup

\end{document}